\pdfoutput=1
\documentclass[prd,aps,preprint,amsmath,nofootinbib,amssymb,eqsecnum,showkeys,tightenlines]{revtex4-1}

\usepackage{amsfonts}
\usepackage{multirow}
\usepackage{caption}
\usepackage{subcaption}
\usepackage{amsmath}
\usepackage{slashed} 
\usepackage{verbatim,graphics,graphicx,color,slashed}

\usepackage[usenames,dvipsnames]{xcolor}

\def\ra{\rightarrow}

\def\wt{\tilde}

\def\ld{\lambda}
\def\f{\frac}
\newcommand{\be}{\begin{equation}}
\newcommand{\ee}{\end{equation}}
\newcommand{\bea}{\begin{eqnarray}}
\newcommand{\eea}{\end{eqnarray}}
\newcommand{\ba}{\begin{array}}
\newcommand{\ea}{\end{array}}

\long\def\symbolfootnote[#1]#2{\begingroup%
\def\thefootnote{\fnsymbol{footnote}}\footnote[#1]{#2}\endgroup}

\newcommand{\beq}{\begin{equation}}
\newcommand{\eeq}{\end{equation}}

\usepackage{ulem} 
\usepackage[colorlinks=true,
            linkcolor=red,
            urlcolor=blue,
            citecolor=blue]{hyperref}

\begin{document}

\title{Uncover Compressed Supersymmetry via Boosted Bosons from the Heavier Stop/Sbottom}

\author{Zhaofeng Kang}
\email{zhaofengkang@gmail.com}
\affiliation{School of Physics, Korea Institute for Advanced Study, Seoul 130-722, Korea}

\author{Jinmian Li}
\email{jmli@kias.re.kr}
\affiliation{School of Physics, Korea Institute for Advanced Study, Seoul 130-722, Korea}

\author{Mengchao Zhang}
\email{mczhang@ibs.re.kr}
\affiliation{Center for Theoretical Physics of the Universe, Institute for Basic Science (IBS), Daejeon, 34051, Korea}

\begin{abstract}

A light stop around the weak scale is a hopeful messenger of natural supersymmetry (SUSY), but it has not shown up at the current stage of LHC. Such a situation raises the question of the fate of natural SUSY. Actually, a relatively light stop can easily be hidden in a compressed spectra such as mild mass degeneracy between stop and neutralino plus top quark. Searching for such a stop at the LHC is a challenge. On the other hand, in terms of the argument of natural SUSY, other members in the stop sector, including a heavier stop $\wt t_2$ and lighter sbottom $\wt b_1$ (both assumed to be left-handed-like), are also supposed to be relatively light and therefore searching for them would provide an alternative method to probe natural SUSY with a compressed spectra. In this paper we consider quasi-natural SUSY which tolerates relatively heavy colored partners near the TeV scale, with a moderately large mass gap between the heavier members and the lightest stop. Then $W/Z/h$ as companions of $\wt t_2$ and $\wt b_1$ decaying into $\wt t_1$ generically are well boosted, and they, along with other visible particles from $\wt t_1$ decay, are a good probe to study compressed SUSY. We find that the resulting search strategy with boosted bosons can have better sensitivity than those utilizing multi-leptons.

\end{abstract}
 
\maketitle

\section{Introduction}

Supersymmetry (SUSY), devised to elegantly solve the gauge hierarchy problem, used to and will provide the major impetus for building high energy colliders, such as the LHC. Guided by the naturalness argument~\cite{Barbieri:1987fn,Kitano:2006gv,Giudice:2006sn,Papucci:2011wy},  stops, among a bunch of new particles predicted by SUSY, should be light and therefore, along with their color charges, may furnish the first ``smoking-gun" signature for SUSY at the LHC. 

Nevertheless, confirmatory hints for light stops at the LHC are absent so far. Considering that the LHC is now already running at the CM energy $\sqrt{s}=13$ TeV, the null results arouse concerns about the existence of low energy SUSY, or more concretely light stop below the TeV scale. Actually, the current LHC search strategies~\cite{Aad:2015pfx,Aaboud:2016tnv,ATLAS:2016xcm,ATLAS:2016ljb,ATLAS:2016jaa,CMS:2016vew,CMS:2016hxa,CMS:2016inz} still leave a wide room for a relatively light stop ($\sim 500$ GeV), provided that a very large missing energy from stop decay is not present, say due to a compressed spectrum. Such a spectrum is characterized by very close mass between stop $\wt t$ and the lightest sparticle (LSP)~\cite{Carena:2008mj,Bornhauser:2010mw,Ajaib:2011hs,Han:2012fw,Drees:2012dd,Cao:2012rz,Yu:2012kj,Krizka:2012ah,Delgado:2012eu,Han:2013kga,Czakon:2014fka,Belanger:2013oka,Dutta:2013gga,Grober:2014aha,Hikasa:2015lma,An:2015uwa,Belanger:2015vwa,Kobakhidze:2015scd,Ferretti:2015dea,Macaluso:2015wja,Goncalves:2016nil,Cheng:2016mcw,Kaufman:2015nda,Duan:2016vpp,Han:2016xet,Goncalves:2016tft}, or more loosely speaking $m_{\wt t}\sim m_t+m_{\rm LSP}$ and $m_{\wt t}\sim m_b+m_W+m_{\rm LSP}$.~\footnote{In the case of a sneutrino LSP, the compressed spectrum may require $m_{\wt t_1}\approx m_t+\ell+m_{\wt \nu_1}$~\cite{Guo:2013asa}, where three-body can help to soften missing energy even the compression is only mild.} Despite allowing for a natural low energy supersymmetry, it is challenging to uncover such a stop at the LHC. Nevertheless, if naturalness is reliable, the heavier stop and the lighter sbottom should not lie far above $m_{\wt t_1}$, thus being detectable. 
This motivates the searches for the signal of $\tilde{t}_2$ pair production with $\tilde{t}_2 \to\tilde{t}_1  h/Z $ decay at the LHC run-I~\cite{Aad:2014mha,Khachatryan:2014doa}. Moreover, through searching for the final state of multi-leptons and/or multi-b-jets~\cite{Guo:2013iij}, the heavier stop/sbottom with masses below $\sim 1$ TeV decaying into $\tilde{t}_1$ and heavy bosons are found to be detectable at the 13/14 TeV LHC with an integrated luminosity of $\mathcal{O}(100)~\text{fb}^{-1}$~\cite{Beuria:2015mta,Cheng:2016npb,An:2016nlb,Pierce:2016nwg}.

On the other hand, boosted objects such as a boosted top quark, vector bosons and the Higgs boson being new physics signatures have been receiving increasing experimental attention~\cite{Butterworth:2008iy,Kaplan:2008ie,Plehn:2009rk,Cui:2010km}, where the new physics scale is pushed into the higher and higher region. The substructures of these boosted objects furnish a powerful tool to distinguish the signatures from the huge QCD backgrounds. Taking into account that they (top etc.) dominantly decay into hadrons, the substructure approach may be more efficient than the searching approach utilizing their leptonic final states. This leads us to reconsider the strategy of searching for the heavier stop/sbottom in the compressed SUSY scenario. If there is a relatively large mass splitting between the heavier stop/sbottom ($\tilde{t}_2 / \tilde{b}_1$) and the lighter stop ($\tilde{t}_1$), the $h/Z/W$ boson in the decay chain $\tilde{t}_2 \to h/Z \tilde{t}_1$ and $\tilde{b}_1 \to W \tilde{t}_1$ will be quite energetic. Hence, hunting for $\tilde{t}_2 / \tilde{b}_1$ by tagging these boosted bosons may be a promising way. It was already tried in an earlier paper~\cite{Ghosh:2013qga}, which employed the boosted boson tag technique to probe the highly mixed stop sector  and obtained a satisfactory sensitivity for $m_{\tilde{t}_2} \sim 1$ TeV and $m_{\tilde{t}_1} \sim 400$ GeV. But this study focused on the case of degeneracy between $\wt t_1$ and the LSP, with $m_{\tilde{t}_1} - m_{\tilde{\chi}^0_1} \sim  \mathcal{O}(10)$ GeV, which requires the flavor-violating decay $\tilde{t}_1\to c \tilde{\chi}^0_1$ and renders $\wt t_1$ invisible. Whereas for a moderately compressed spectrum considered in this paper, additional visible particles from $\wt t_1$ (flavor conserving) decay are available.

So, in this paper we consider a (simplified) quasi-natural pattern of low energy supersymmetry where the lighter stop of a few hundred GeV is right-handed stop like and lives in the compressed regions, due to its close mass with bino or Higgsinos; whereas states in the doublet $\wt Q_3$ are around the TeV scale. Thus, the characteristic signatures of this model contain fairly boosted bosons from decays $\wt Q_3\ra \wt t_1+W/Z/h$. To demonstrate the prospects of those signatures at the LHC, we choose four benchmark points corresponding to four possible decay modes of $\tilde{t}_1$: (1) $\tilde{t}_1 \to b \tilde{\chi}^\pm_1$; (2) $\tilde{t}_1 \to b f f \tilde{\chi}^0_1$; (3) $\tilde{t}_1 \to b W \tilde{\chi}^0_1$; (4) $\tilde{t}_1 \to t \tilde{\chi}^0_1$, which produce extra detectable b-jets and leptons as well as missing transverse energy (MET). Therefore, boosted bosons plus MET, associated with $b$-jets/leptons constitute the smoking-gun signature for such a compressed SUSY. By adopting the boosted decision tree (BDT) method for signal and background discrimination, we find that the resulting search strategy with boosted bosons can have better sensitivity than those utilizing multi-leptons.

The paper is organized as the following. In Section~\ref{sec:model} we establish the quasi-natural SUSY which can hide the lighter stop involving the minimal degrees of freedom and demonstrate the distribution of $\wt t_2$ and $\wt b_1$ decays in the MSSM. In Section~\ref{sec:collider} we detail the signal and background analysis at the LHC. Discussions and conclusions are presented in the final section.

\section{Quasi-natural supersymmetry}
\label{sec:model}

In this section we will present the quasi-natural model with minimal field content and analyze the decay modes of the heavier stop and sbottom, in particular the bosonic modes, analytically and numerically. Accordingly, benchmark points are selected out.

\subsection{A minimal setup}

Asides from a light stop sector, naturalness arguments in general favor a weak scale $\mu$-term thus light Higgsinos. On the other hand, considering the SUSY status after the discovery of a relatively heavy SM-like Higgs boson but no hints for light stops, we may have to abandon the ideal naturalness criterion and tolerate fine-tuning to some degree, says 1\% or even worse~\cite{Kang:2012sy,Cao:2012fz}. Such a situation inspires us to consider a quasi-natural SUSY involves a minimal set of particles that accommodate a light stop $\wt t_1$ with or without weak scale Higgsinos; other superpartners including $\wt b_R$ and winos, are simply assumed to decouple for simplicity. The resulting Lagrangian most relevant to our discussions derived from the flavor basis is (We just schematically list the terms.)~\footnote{We do not consider the way to obtain a 125 GeV SM-like Higgs boson mass in this paper; other sources of Higgs mass should be introduced, otherwise the stop sector will be pushed into the multi-TeV region and almost inaccessible at the LHC.}  
\begin{align}\label{QNS:model}
-&{\cal L}_{QNS}=m_{\wt Q_3}^2|\wt Q_3|^2+\f{m_{\tilde{B}}}{2} \wt B\wt B+m_{\wt t_R}^2|\wt t_R|^2+\left( h_tA_t\wt Q_3 H_u t_R^\dagger+|D_\mu\wt Q_3|^2\right)\cr
&+i\sqrt{2}\left(\f{g_Y}{6}\wt Q_3^\dagger \wt B t_L-\f{2g_Y}{3}\wt t_R^\dagger \wt B t_R\right)+\left( \mu \wt H_u\wt H_d+h_t \wt Q_3 \wt H_u t_R^\dagger +h_b 
\wt Q_3\wt H_d b_R^\dagger\right)+h.c.,
\end{align}
where $D_\mu=\partial-i\f{g_2}{\sqrt{2}}\left(T^+W_\mu^++T^-W_\mu^-\right)-i\f{g_2}{\cos\theta_W}Z_\mu\left(T^3-\sin^2\theta_WQ\right)+...$ with $\theta_W$ the weak mixing angle. In the second line, terms in the first and second brackets may be irrelevant if $\wt B$ and $\mu$ are much heavier than all other particles therein, respectively. For simplicity, we will consider that either bino or Higgsino is light and provide the LSP. Although a large $A_t$ is not necessarily required in this setup, we will see that it is crucial viewing from collider searches; besides, recalling the difficulty in achieving a relatively heavy SM-like Higgs boson in natural SUSY, a large $A_t$, which could really help to radiatively enhance Higgs boson mass, is well motivated. A good case in point of such kind of quasi-natural SUSY is the Higgs deflected gauge mediated SUSY-breaking~\cite{Kang:2012ra}. 

A compressed superpartner spectra could make $\wt t_1$ hard to detect. If the mass degeneracy between $\wt t_1$ and LSP is mild and $\wt t_1\ra t+\wt \chi^0_1$ or $b+\wt\chi_1^\pm$ is kinematically accessible, they will become the main decay modes of $\wt t_1$. If degeneracy becomes severer, the above channels are closed and $\wt t_1$ will dominantly three (four)-body decays into $bW^{(*)}\wt \chi^0_1$, assuming that the flavor changing decay $\wt t_1\ra c \wt \chi^0_1$, which strongly depends on the unknown flavor structure of squarks, is negligible. As a matter of fact, the four-body decay case is particularly well motivated after identifying bino as the dark matter candidate: Bino is a gauge singlet, so, in order to reduce its relic density during the freeze-out era,  usually coannihilation with a nearly degenerate stop is necessary;  for a sub TeV bino DM, a fairly small mass difference $m_{\wt t_1}-m_{\wt\chi_1^0}\sim$ 30GeV is needed~\cite{Ellis:2001nx}.

To hide a light $\wt t_1$ at the current LHC, it is better to let $\wt t_1$ dominantly reside in $\wt t_R$; otherwise, the accompanied $\wt b_L$ which has close mass with $\wt t_1\approx \wt t_L$ would have been uncovered via $\wt b_L\ra \wt \chi^0_1 b$ except for highly degeneracy between $\wt t_1$ and $\wt \chi^0_1$, a case has been extensively discussed before~\cite{An:2016nlb}. Moreover, in this paper we focus on that the doublet $\wt Q_3=(\wt b_L, \wt t_L)$ are considerably heavier than $\wt t_R\approx \wt t_1$, and therefore, by tagging the boosted bosons from $\wt Q_3$ decaying into $\wt t_1$, they may show more promising prospect at the LHC than $\wt t_1$, which is somewhat hidden as before. On the contrary, $\wt Q_3$ having similar mass to $\wt t_1$ may be hard to discover, because their decay final states typically are soft. In this sense the heavier stop/sbottom may instead provide the smoking gun for (quasi-)natural SUSY.

\subsection{Bosonic decay modes of $\wt t_2$ and $\wt b_1$: Roles of a large $A_t$}
\label{sec:decay}

In this subsection, we examine the bosonic decay modes of $\wt t_2$ and $\wt b_1$ and see the conditions which make them be the dominant modes. These decays do not depend on the nature of LSP. Concretely, their decay widths are given by~\cite{bartl1994squark}
\begin{align}\label{}
\Gamma(\wt b_1 \ra \wt t_1 W)&\approx\f{g_2^2\cos^2\theta_{\wt
t}}{32\pi}\f{m_{\wt b_1 }^3}{m_W^2}\ld^{3/2}(m^2_{\wt b_1},m^2_{\wt
t_1},m^2_W),\cr
\Gamma(\wt t_2\ra \wt
t_1Z)&\approx\f{g_2^2}{\cos^2\theta_W}\f{\sin^22\theta_{\wt
t}}{256\pi}\f{m_{\wt t_2}^3}{m_Z^2}\ld^{3/2}(m^2_{\wt t_2}, m^2_{\wt
t_1}, m^2_Z),\cr
\Gamma(\wt t_2\ra \wt
t_1h)&\approx\f{g_2^2\cos^22\theta_{\wt t}}{64\pi}\f{m_t^2}{m_W^2}\f{A_t^2}{m_{\wt
t_2}} \ld^{1/2}(m^2_{\wt t_2}, m^2_{\wt
t_1}, m^2_h),
\end{align}
with $\ld(a,b,c)=[1-(b+c)/a]^2-4bc/a^2\approx 1$. We have taken $h\sim {\rm Re}(H_u^0)$ to get the last expression. Here $\wt \theta_t$ is the mixing angle between the left- and right-handed stops, defined through  
\begin{align}\label{}
\wt t_{L}=\cos{\theta_{\wt t}}\wt t_{1}-\sin\theta_{\wt t}\wt
t_2,\quad \wt t_{R}=\sin{\theta_{\wt t}}\wt t_{1}+\cos\theta_{\wt
t}\wt t_2,
\end{align}
and $\tan2\theta_{\wt t}=2X_t m_t/(m_{\wt Q_3}^2-m^2_{\wt t_R})$ with $X_t=A_t+\mu/\tan\beta$. If $\wt t_1$ is very $\wt t_R$-like, one will have $\wt\theta_t\ra\pi/2$ and consequently all the bosonic modes except for $\wt t_2\ra\wt t_1 h$ will be highly suppressed. A large $A_t$ is thus indispensable: It does not only generate sizable LR stop mixing but also directly enhance $\wt t_2\ra \wt t_1h$.~\footnote{In Ref.~\cite{Kang:2015nga}, assuming that $\wt t_1$ behaves like a pure missing energy at the LHC, says in the highly compressed limit, a new search approach based on boosted di-Higgs plus missing energy was proposed.} In practice, we do not need a fairly sizable $\theta_{\wt t}$ because the (longitudinal) $W/Z$ modes are enhanced by a factor like $\left( m_{\wt t_2}/m_Z\right)^2\sim{\cal O}(10^2)$ for a TeV scale $m_{\wt Q_3}$, which could easily compensate the mild suppression from the small mixing.

Now we analyze their heavy quark decay modes based on the quasi-natural SUSY~Eq.~(\ref{QNS:model}), which are sensitivity to the LSP components. In the most general cases, the decay widths take the forms of~\cite{bartl1994squark}
\begin{align}\label{}
\Gamma(\wt q_i\ra q\wt\chi^0_k)&=\f{g_2^2}{16\pi m_{\wt q_i}}\left[ \left((h^q_{ik})^2+(f^q_{ik})^2\right)\left(m_{\wt q_i}^2-m_q^2-m_{\wt \chi^0_k}^2\right)-4h^q_{ik}f^q_{ik}m_q m_{\wt \chi^0_k}
   \right],\\
   \Gamma(\wt q_i\ra q'\wt\chi^\pm_k)&=\f{g_2^2}{16\pi m_{\wt q_i}}\left[ \left((l^{\wt q}_{ik})^2+(k^{\wt q}_{ik})^2\right)\left(m_{\wt q_i}^2-m_{q'}^2-m_{\wt \chi^0_k}^2\right)-4l^{\wt q}_{ik}k^{\wt q}_{ik}m_{q'} m_{\wt \chi^\pm_k}
   \right] . 
\end{align}
where $\wt q_i$ denote for $\wt t_{1, 2}$ and $\wt b_{1}$. The matrices $h^q_{ik}$ etc., encode couplings between quark and squark, neutralinos; in the following we will give their concrete expressions in the Higgsino- and bino-LSP limites. 

We first consider the Higgsinos are light while bino can be dropped; moreover, we will use the strip $m_b+|\mu|< m_{\wt t_1}\lesssim m_t+|\mu|$ to hide $\wt t_1$. In the limit of a left-handed sbottom while right-handed light stop, namely $\theta_{\wt t}\ra\pi/2$, and as well a Higgsino LSP (actually two with almost degenerate masses involved) one obtains
\begin{align}\label{}
h_{21}^t&\approx h_{22}^t\approx 0, \quad |f_{21}^t|\approx  |f_{22}^t|\approx \f{m_t}{2m_W}\sin\theta_{\wt t},\cr
|h_{11}^b|&\approx |h_{12}^b|\approx  \f{m_b}{2m_W}\tan\beta, \quad f_{11}^b\approx  f_{12}^b\approx 0,\\
l_{21}^{\wt t}&\approx 0 ,\quad   |k_{21}^{\wt t}| \approx\f{m_b}{\sqrt{2}m_W\cos\beta}\sin\theta_{\wt t}, \cr
 l_{11}^{\wt b}&\approx 0 ,\quad   |k_{11}^{\wt b}| \approx\f{m_t}{\sqrt{2}m_W\sin\beta},
\end{align}
where a relatively large $\tan\beta$ at least a few is assumed. Next we move to the other case where Higgsinos are decoupled and bino is the LSP. In this case a lighter $\wt t_1$ is allowed if its mass does not significantly exceed $m_t+m_{\wt\chi_1^0}$. Now the couplings are reduced to
\begin{align}\label{}
|h_{21}^t|&\approx \f{2\sqrt{2}}{3}\sin\theta_W\sin\theta_{\wt t}, \quad  |f_{11}^b|\approx  \f{\sqrt{2}}{3}\sin\theta_W\sin\theta_{\wt t}.
\end{align}
All others are suppressed by small mixing angles thus of no importance. Moreover, since the charginos are decoupled, here we do not need to consider $l^{\wt b}_{1j}$, etc.

We would like to stress that, in the bino-LSP case the decay modes of $\wt t_2$ and $\wt b_1$ into the heavy flavors, such as $\wt t_2\ra t+\wt\chi_1^0$ and $\wt b_1\ra b+\wt\chi_1^0$, are substantially suppressed, because now they come from hypercharge gauge interactions rather than the $y_t$-Yukawa interaction as in the light Higgsino case. One can clearly see this situation from the Fig.~\ref{brs}, which shows that those branching ratios typically are below ${\cal O}(1\%)$ in the bino LSP scenario. Such a situation makes good for the more boosted bosons from $\wt t_2/\wt b_1$ decay. But even in the Higgisno LSP case,  for a heavier $m_{\wt Q_3}$ with a large $A_t$ coupling, these bosonic modes generically have quite sizable branching ratios, $\gtrsim{\cal O}(10\%)$, by virtue of the significant Goldstone enhancement factor stressed before. In particular, $\wt b_1$, which has less decay modes than $\wt t_2$, almost dominantly decays into $W$ plus $\wt t_1$ in both cases. It is can be understood from the estimation (in the Higgsino-LSP limit): 
\begin{align}\label{}
\f{\Gamma(\wt b_1\ra W\wt\chi^\pm_1)}{\Gamma(\wt b_1\ra b \wt\chi^0_1)}\simeq \left(\f{m_{\wt b_1}}{m_t}\right)^2\cos^2\theta_{\wt t}\gtrsim \f{A_t^2}{m_{\wt Q_3}^2}.
\end{align}

In the next section, we will choose several benchmarks points to embody the above possible scenarios for quasi-natural SUSY.

\subsection{Decay patterns in quasi natural SUSY: Scanning results and Benchmark points}

For concreteness, we implement quasi-natural SUSY in the minimal supersymmetric SM (MSSM). There are totally 5 parameters of interests in each scenario with either decoupled bino or Higgsino.  We use the Suspect2~\cite{Djouadi:2002ze} and SUSY-HIT~\cite{Djouadi:2007aa} to calculate the mass spectra and the decay branching ratios of stops and sbottom.  The parameter scan is performed in the following range:
\begin{align} \label{prange}
 m_{\wt Q_3} \in [700, 1200] ~\text{GeV}, ~& m_{\wt U_3} \in [500,700] ~\text{GeV}, \cr 
 |A_t| \in [1,3] ~\text{TeV}, ~& \tan \beta \in [3,50], \cr
 \mu \in [300,700] ~\text{GeV} ~(M_1=2~\text{TeV}) \text{~or~} & M_1 \in [300, 700]~\text{GeV} ~(\mu=2~\text{TeV}).
\end{align}
The rest of the soft mass parameters of the MSSM are set to 2 TeV so those sparticles are decoupled from the mass spectrum. The choice of the above parameter pattern is motivated by non-detections of any stop/sbottom signals at the current stage of LHC~\cite{Aad:2015pfx,Aaboud:2016tnv,ATLAS:2016xcm,ATLAS:2016ljb,ATLAS:2016jaa,CMS:2016vew,CMS:2016hxa,CMS:2016inz}; the resulting spectra still allows a light stop with mass $\sim 500$ GeV if the LSP is relatively heavy ($m_{\rm LSP} \gtrsim 300$ GeV). Moreover, we require the mass of the heavier stop and the sbottom to be around the TeV scale to produce relatively boosted bosons in their decay while still having sizable production rates for discovery in the near future. We note that the searches for heavier stop at the LHC run-I~\cite{Aad:2014mha,Khachatryan:2014doa} are only able to exclude models with $m_{\tilde{t}_2} \lesssim 600$ GeV. As we have discussed in Sec.~\ref{sec:decay}, a sizable $|A_t|$ is needed to enhance ${\rm Br}(\wt t_2 \ra h\wt t_1)$ and ${\rm Br}(\wt t_2 \ra Z\wt t_1)$, so a lower limit of $|A_t|$ is set to improve the scanning efficiency.  
Since we are expecting new contributions other from the stop in MSSM to the Higgs boson mass, the lighter CP-even Higgs boson ($H_1 \equiv h$) mass is set to 125 GeV manually when calculating the decay branching ratios. 
The heavier CP-even Higgs ($H_2$) is decoupled by setting $m_A=2$ TeV.  

\begin{figure}[th]
\begin{center}
\includegraphics[width=0.49\textwidth]{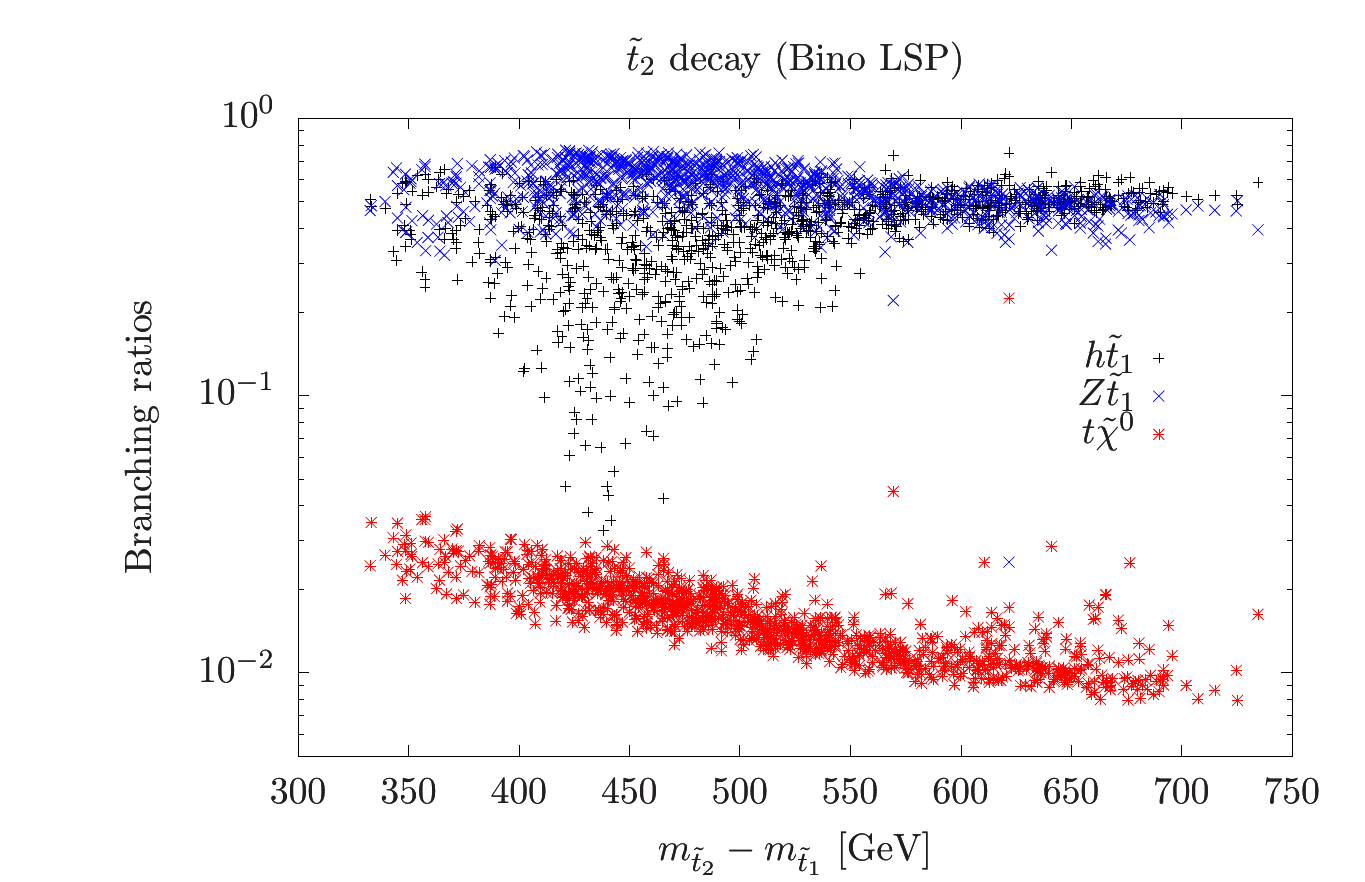}
\includegraphics[width=0.49\textwidth]{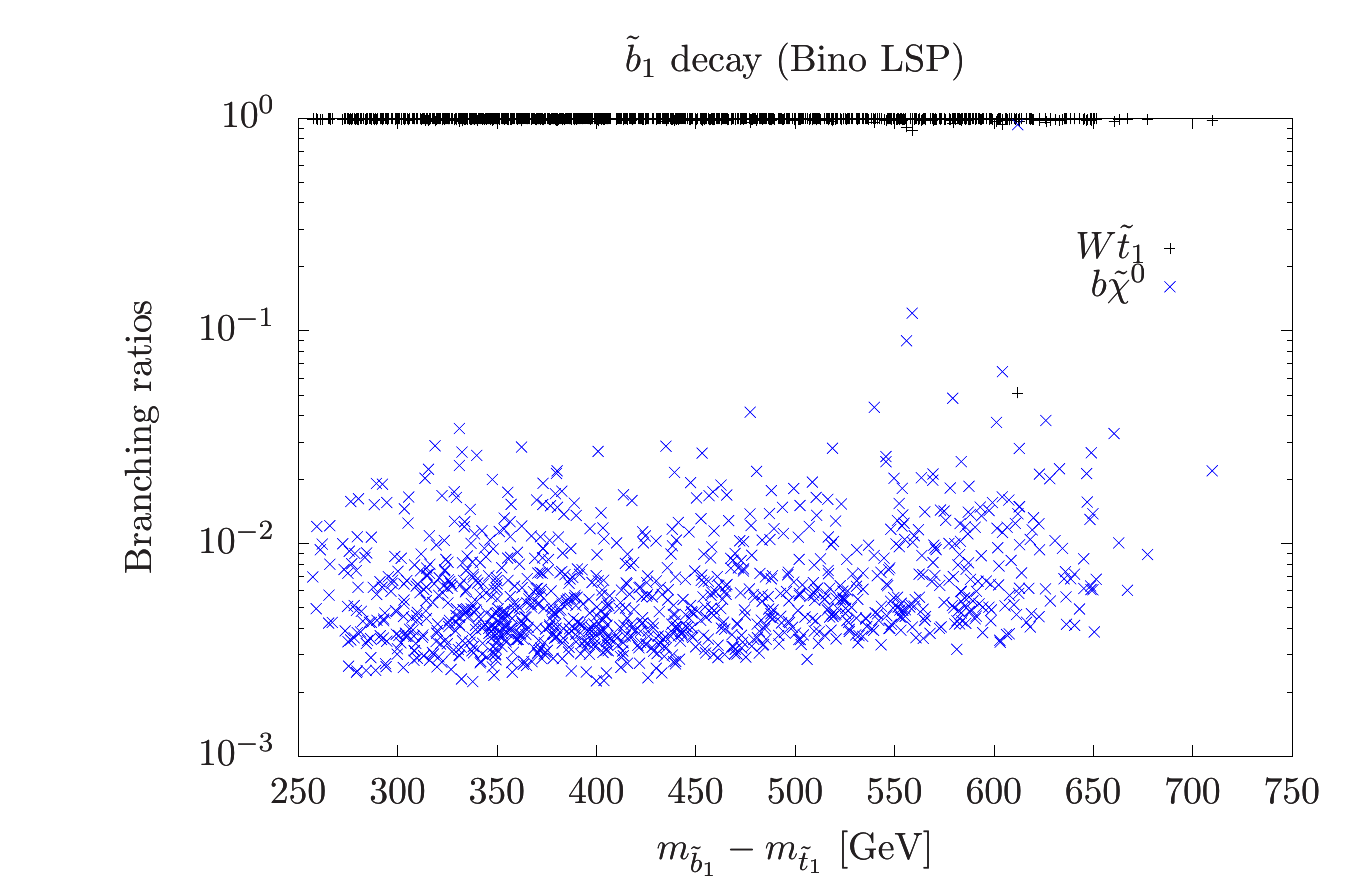} \\
\includegraphics[width=0.49\textwidth]{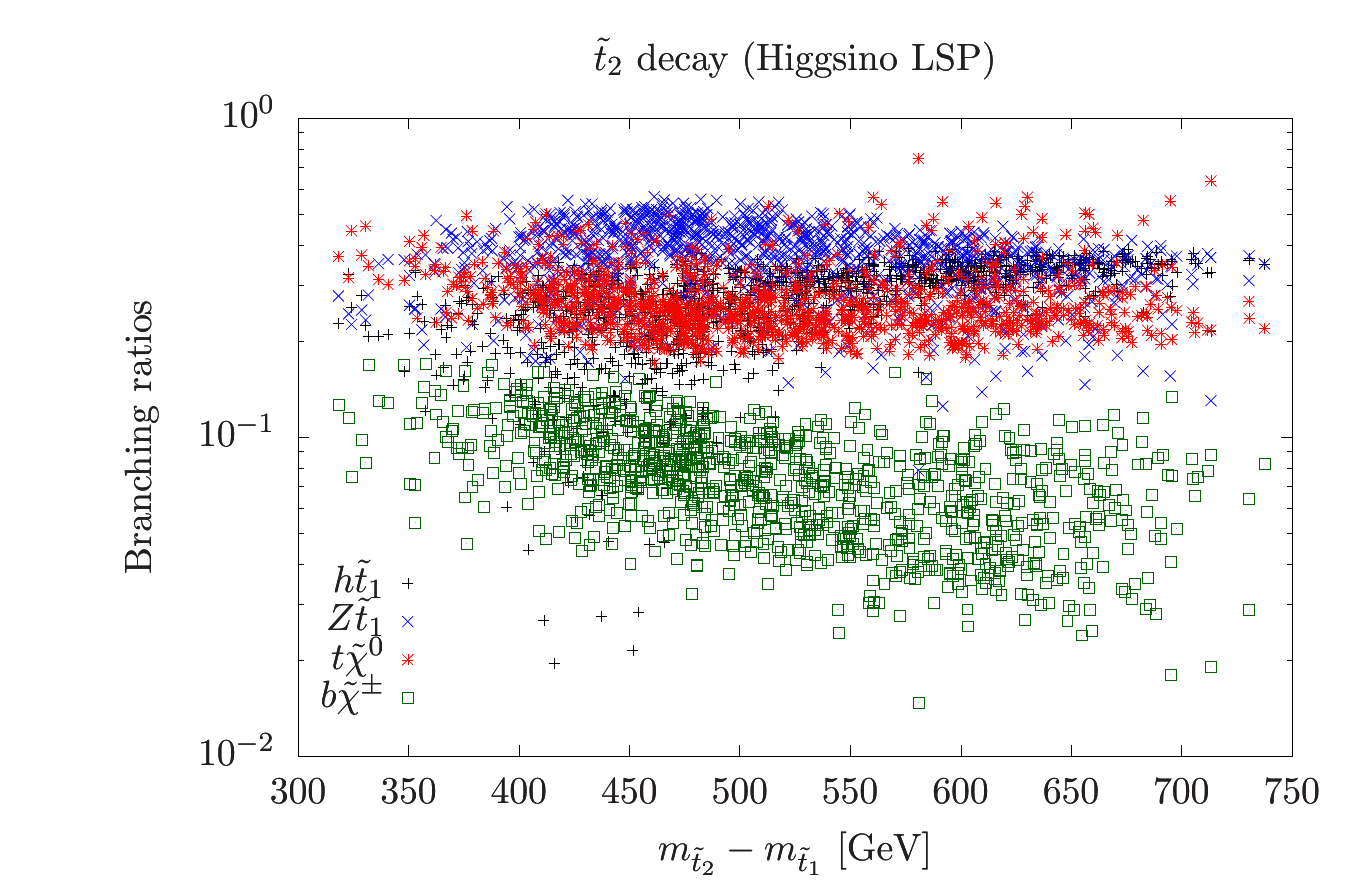}
\includegraphics[width=0.49\textwidth]{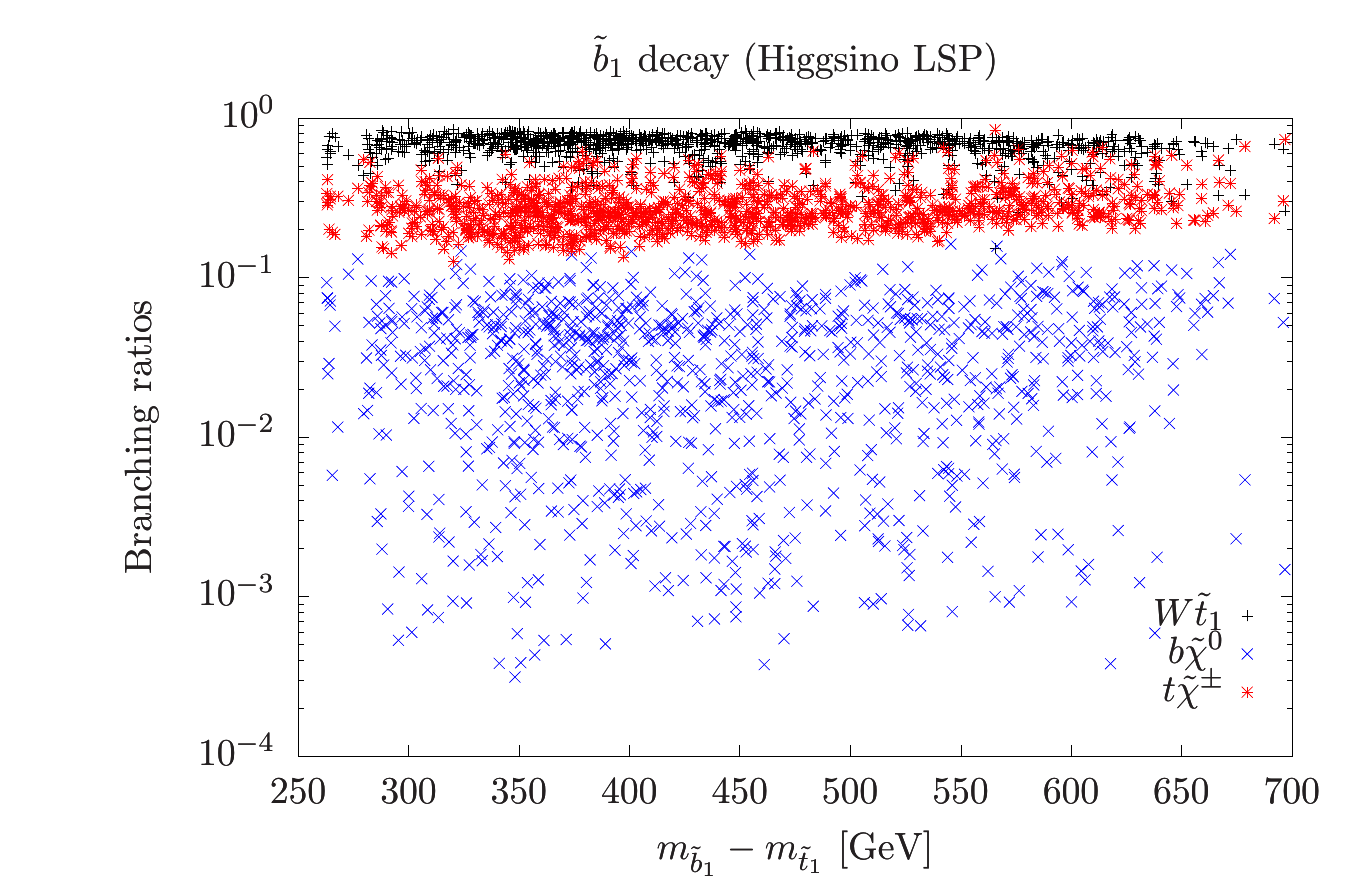}
\end{center}
\caption{\label{brs} $\tilde{t}_2$ (left panels) and $\tilde{b}_1$ (right panels) decay branching ratios for Bino LSP (upper panel) and Higgsino LSP (lower panels).  }
\end{figure}

In Fig.~\ref{brs}, we plot the decay branching ratios of the heavier stop ($\tilde{t}_2$) and the lighter sbottom ($\tilde{b}_1$) for either bino LSP or Higgsino LSP.  In the upper panels where bino is the LSP, we can see that the bosonic modes dominate the stop/sbttom decay in the fully parameter space while the branching fractions of $\tilde{t}_2 \to t \tilde{\chi}^0_1$/$\tilde{b}_1 \to b \tilde{\chi}^0_1$ modes typically are two order of magnitude smaller. The situation changes when Higgsino is the LSP. In the lower panels, decay widths of $\tilde{t}_2 \to t \tilde{\chi}^0_1$/$\tilde{b}_1 \to b \tilde{\chi}^0_1$ which are enhanced by the larger top quark Yukawa coupling become comparable to that of the bosonic modes. Moreover, there will be new decay modes opening due to the charged Higgsino in the final state, i.e., $\tilde{t}_2 \to b \tilde{\chi}^\pm_1$ and $\tilde{b}_1 \to t \tilde{\chi}^\pm_1$ whose branching fractions are also sizable. Nevertheless, we can observe that the bosonic mode is still one of the dominant decay mode for both $\tilde{t}_2$ and $\tilde{b}_1$.

In terms of the scanning results, eight benchmark points are chosen to illustrate the model details in Tab.~\ref{bps}, which are featured by the different decay modes of the lighter stop, as well as two choices of the $\tilde{t}_2/ \tilde{b}_1$ masses, that is $m_{\tilde{t}_2/ \tilde{b}_1} \sim 800$ GeV and $m_{\tilde{t}_2/ \tilde{b}_1} \sim 1000$ GeV respectively. These differences will be used to label each benchmark point in the following discussions, e.g.,T1BC (800) corresponding to the one which has $m_{\wt t_2/\wt b_1}\sim 800$ GeV along with a lighter stop mainly decaying into $b+\wt \chi_1^0$.

\begin{table}[htb]
\begin{center}
\begin{tabular}{|c|c|c|c|c||c|c|c|c|} \hline
 & \multicolumn{4} {|c|} { $m_{\tilde{t}_2/ \tilde{b}_1} \sim 800$ GeV} &  \multicolumn{4} {|c|} { $m_{\tilde{t}_2/ \tilde{b}_1} \sim 1000$ GeV}   \\
  & T1BC & T14B & T1BW & T1TN & T1BC & T14B & T1BW & T1TN \\ \hline \hline
  $M_1$ [GeV] & 2000 & 450 & 380 & 340 & 2000 & 429 &  370 & 330    \\ \hline
  $\mu$ [GeV] & 470 & 2000 & 2000 & 2000 &  470 & 2000 & 2000 &  2000  \\ \hline
  $\tan \beta$ & 3.0 & 3.0 & 3.0 & 3.0 &  3.0 & 5.0  & 5.0 & 5.0    \\ \hline
  $m_{\wt Q_3}$ [GeV] & 830 & 870 & 870 & 870 & 1000 & 1000 & 1000 & 1000  \\ \hline
  $m_{\wt U_3}$ [GeV] & 650 & 620 & 620 & 620 &  650 & 630 &  630 & 630 \\ \hline
  $A_t$ [GeV] & 1000 & 1000 & 1000 & 1000 & 1000 & -1000 & -1000 & -1000  \\ \hline \hline
$m_{\tilde{t}_1}$/GeV & 518 & 533 & 533 & 533 & 574 & 517 & 517 & 517 \\ \hline
$m_{\tilde{t}_2}$/GeV & 810  & 826 & 826 & 826 & 993 & 984 & 984 & 984 \\ \hline
$m_{\tilde{b}_1}$/GeV & 774 & 821 & 821 & 821 & 977 & 968 & 968 & 968\\ \hline
$m_{\tilde{\chi}^0_1}$/GeV  & 470 & 454 & 383 & 343 & 471 & 434 & 374 & 334 \\ \hline
$m_{\tilde{\chi}^0_2}$/GeV & 475 & 2000 & 2000  & 2000 & 476 & 2000 & 2000  & 2000  \\ \hline
$m_{\tilde{\chi}^\pm_1}$/GeV & 472 & 2000 & 2000 & 2000 & 472 & 2000 & 2000  & 2000 \\ \hline \hline

Br$(\tilde{t}_2 \to h \tilde{t}_1)$ & 0.163 &  0.399 &  0.393 & 0.389 & 0.204 &0.502 & 0.501 & 0.500  \\  \hline
Br$(\tilde{t}_2 \to Z \tilde{t}_1)$ & 0.330 &  0.526 &  0.518 & 0.514 & 0.219 & 0.486 & 0.485 & 0.484 \\  \hline
Br$(\tilde{b}_1 \to W \tilde{t}_1)$ & 0.621 & 0.963 & 0.954 & 0.949 & 0.431 & 0.988 & 0.987 & 0.986\\  \hline \hline

Br$(\tilde{t}_1 \to b \tilde{\chi}^\pm_1)$ & 1.0 & 0 & 0  & 0 & 1.0 & 0 & 0  & 0 \\ \hline
Br$(\tilde{t}_1 \to bW^{(*)} \tilde{\chi}^0_1)$ & 0 & 1.0 & 1.0 & 0 & 0 & 0.882 & 1.0 & 0   \\ \hline
Br$(\tilde{t}_1 \to t \tilde{\chi}^0_1)$ & 0 & 0 & 0 & 1.0  & 0 & 0 & 0 & 1.0 \\ \hline
\end{tabular}
\caption{\label{bps}Benchmark points for different decay modes of the right-handed dominant $\tilde{t}_1$. The Br$(\tilde{t}_1 \to bW^{(*)} \tilde{\chi}^0_1)$ of T14B (1000) is slightly smaller than one because the flavor changing decay $\tilde{t}_1 \to c \tilde{\chi}^0_1$ is also important here.}
\end{center}
\end{table}

\section{Collider phenomenology}
\label{sec:collider}

\subsection{Search strategy}

For all benchmark points, the left-handed-like $\tilde{t}_2$ and $\tilde{b}_1$ have similar masses, both dominantly decaying into gauge boson/Higgs boson plus the lighter stop $\tilde{t}_1$. The high multiplicity of gauge bosons in the final state will lead to multiple leptons events. Moreover, one would expect two bottom quark jets in the final state if the flavor changing decay of $\tilde{t}_1 \to c \tilde{\chi}^0_1$ is suppressed. Studies~\cite{Aad:2014mha,Khachatryan:2014doa,Beuria:2015mta,Cheng:2016npb,Pierce:2016nwg} have shown that searching for final states with leptons and b-jets provides as a good probe to the light stop sector, especially when $m_{\tilde{t}_1} \sim m_t + m_{\tilde{\chi}^0_1}$. However, for some of our benchmark points, e.g. T14B, the small mass difference between $m_{\tilde{t}_1}$ and $m_{\tilde{\chi}^0_1}$ may render the b-jets/leptons undetectable. 

To see the point more clearly, we generate the parton level events for our benchmark points with MadGraph5~\cite{Alwall:2014hca}, which are passed to Pythia6~\cite{Sjostrand:2006za} for particle decay, parton showering and hadronization. The Delphes3~\cite{deFavereau:2013fsa} with input of default ATLAS detector card is used for simulating detector effects. In this work, we take the b-jet tagging efficiency as 70\% with the other light quark and gluon mis-tagging probability 1\%~\cite{Aad:2015ydr}.

\begin{figure}[t]
\begin{center}
\includegraphics[width=0.47\textwidth]{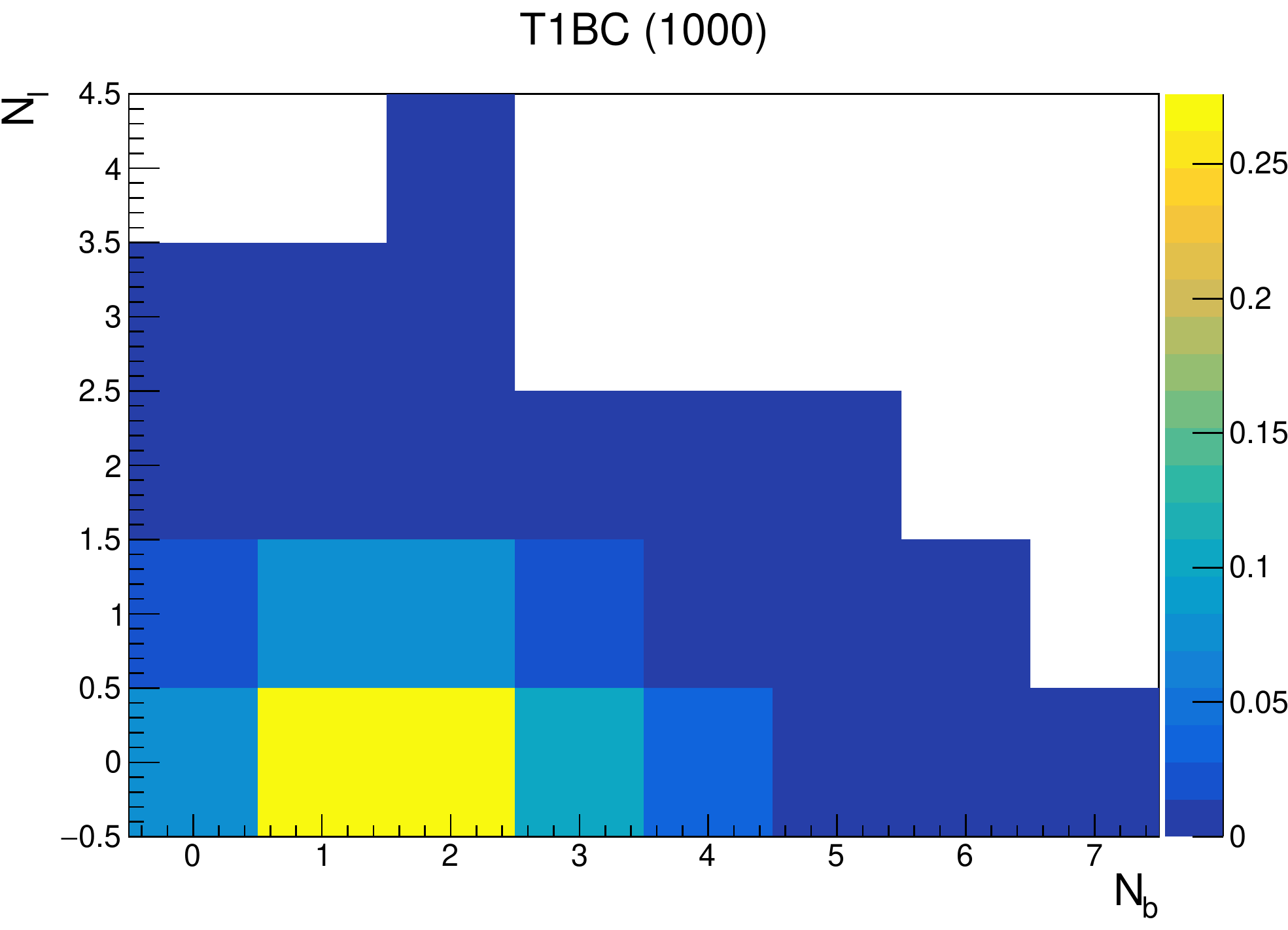}
\includegraphics[width=0.47\textwidth]{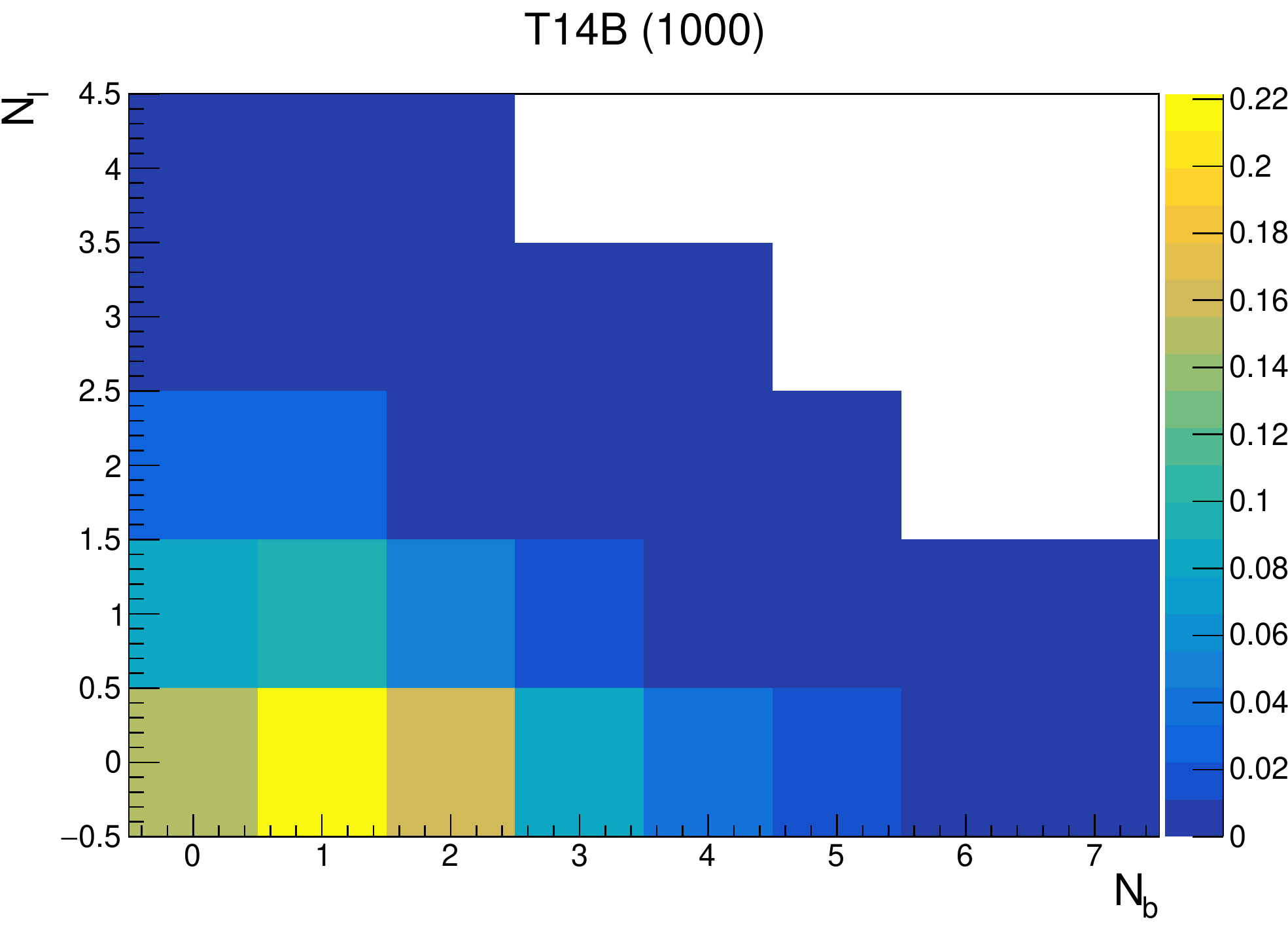} \\
\includegraphics[width=0.47\textwidth]{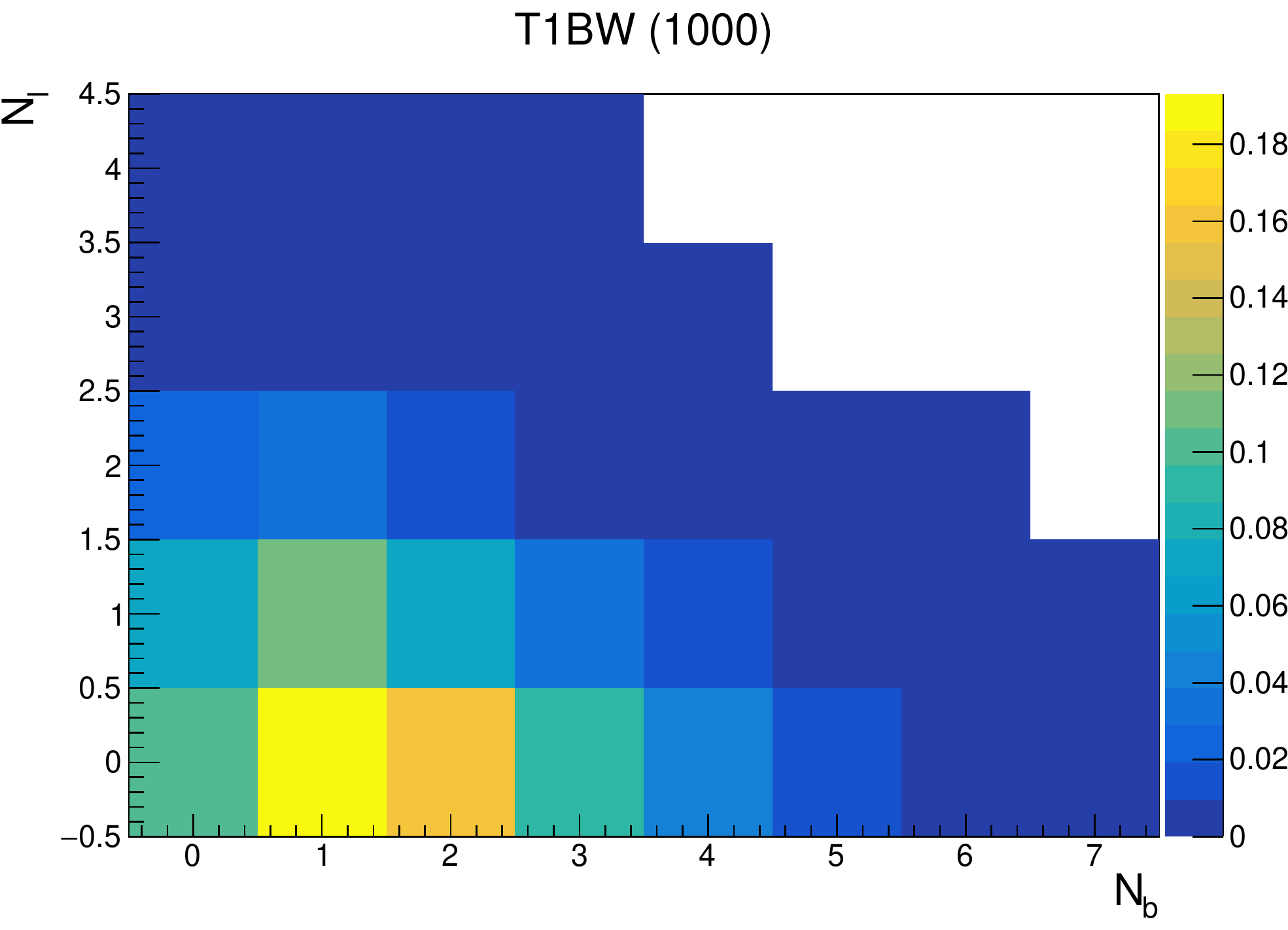} 
\includegraphics[width=0.47\textwidth]{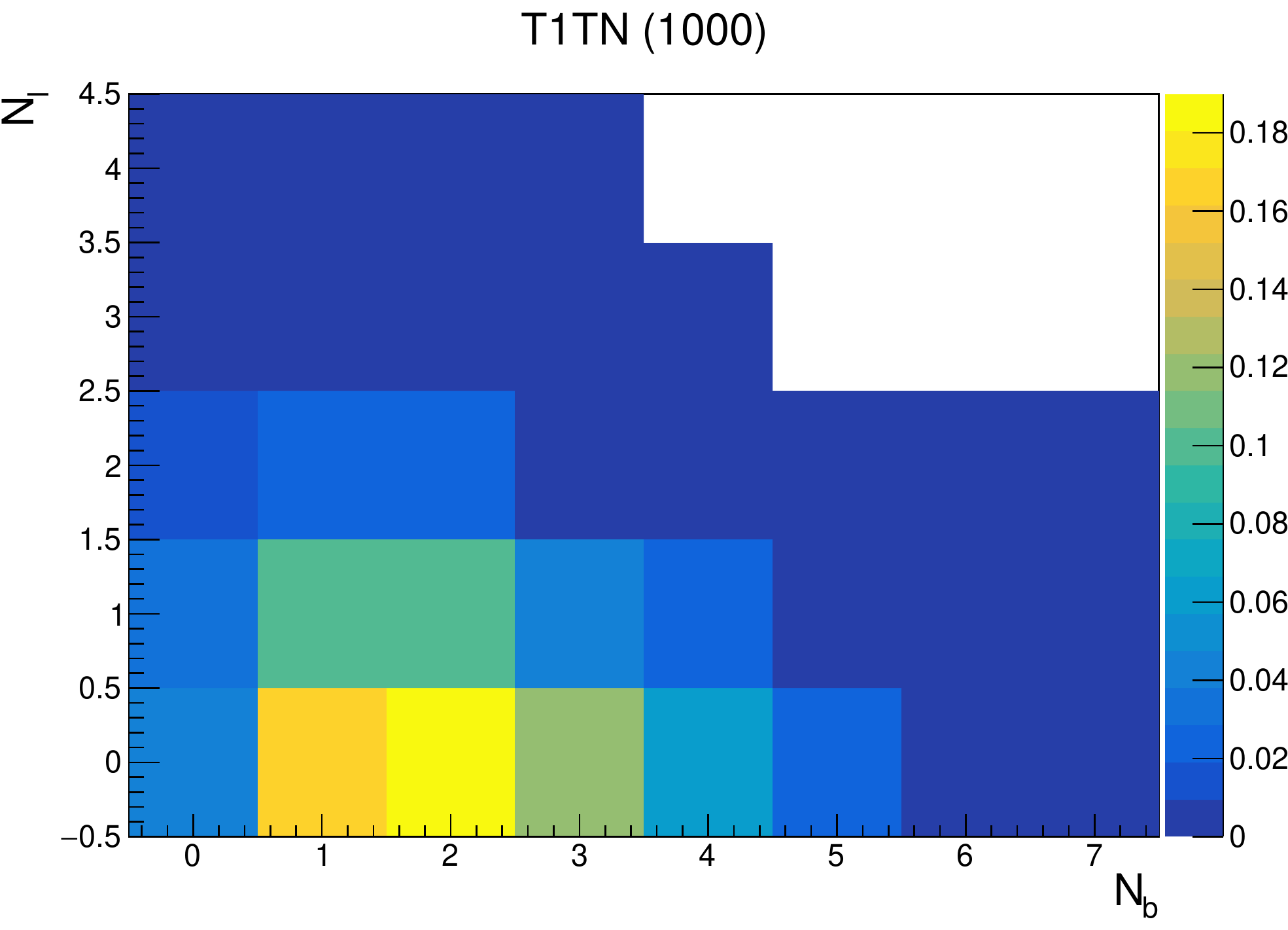}
\end{center}
\caption{\label{nlnb} The number of b-jets $N_b$ and the number of leptons $N_{\ell}$ distributions for our benchmark points with $m_{\tilde{t}_L}$=1 TeV.}
\end{figure}

We consider the signals of both the $\tilde{t}_2$ and $\tilde{b}_1$ pair production with subsequent decays for benchmark points with $m_{\tilde{t}_L}$=1 TeV. The corresponding $N_b$ versus $N_{\ell}$ distributions are given in the Fig.~\ref{nlnb}. It can be seen that even for the benchmark point T1TN in which $\tilde{t}_1$ dominantly decays into $t\tilde{\chi}^0_1$, only around 20\% of the total events contain at least one b-jet and one lepton. The fraction becomes even smaller for other benchmark points owing to the heavier $\tilde{\chi}^0_1$. Events with b-jet multiplicity higher than 2 are originated from $h_{\text{SM}} \to b\bar{b}$. In all cases, we find that the fraction of events with $N_{l} \ge 2$ is at the percent level. Consequently, despite of relatively low backgrounds, searching for final states with multiple leptons is suffering from serious branching ratio suppressions in the signal processes.   

\begin{figure}[t]
\begin{center}
\includegraphics[width=0.47\textwidth]{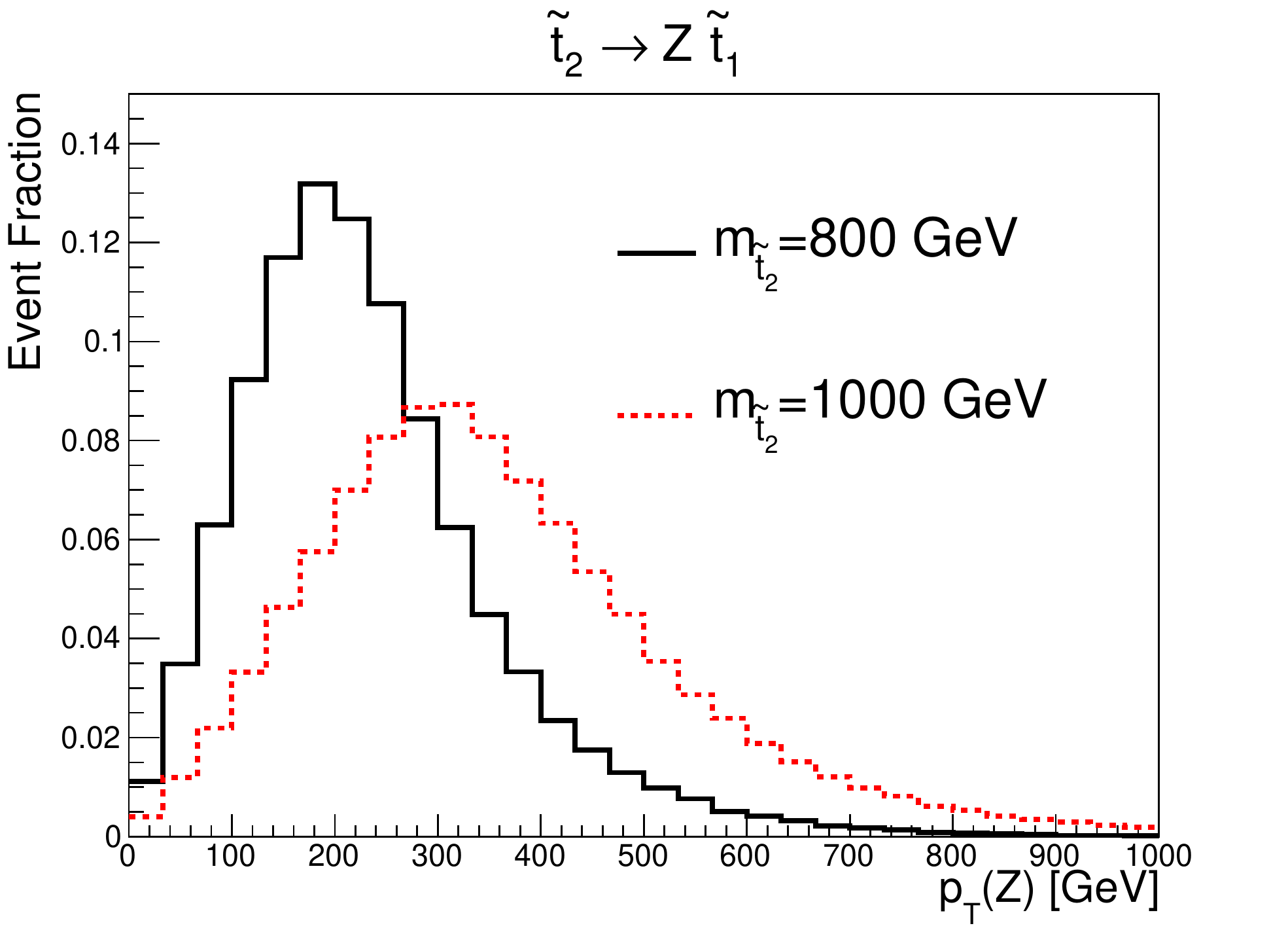}
\includegraphics[width=0.47\textwidth]{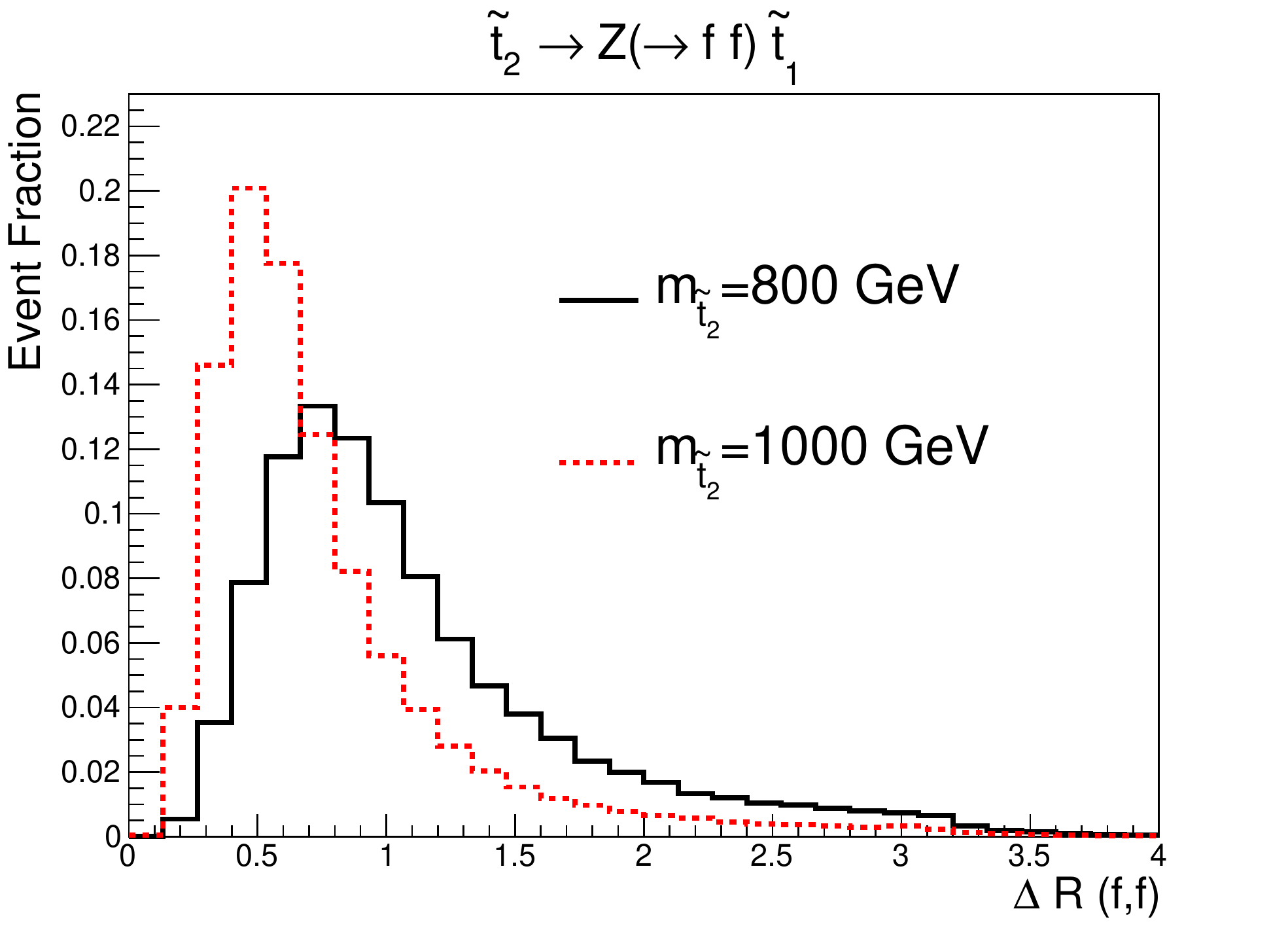}
\end{center}
\caption{\label{pkins} Left: the transverse momentum distribution for the $Z$ boson in $\tilde{t}_2 \to Z \tilde{t}_1$, taking $m_{\tilde{t}_1}=500$ GeV and $m_{\tilde{t}_2}=800/1000$ GeV. Right: the angular distance $\Delta R= \sqrt{(\Delta \phi)^2 + (\Delta \eta)^2}$ between the two fermions from the $Z$ boson decay. }
\end{figure}

On the other hand, signals with hadronic decaying bosons from $\tilde{t}_2/\tilde{b}_1$ decay have much larger production rates. Moreover, some recent developments in the jet substructure analysis~\cite{Butterworth:2008iy,Cui:2010km,Altheimer:2013yza,Rentala:2014bxa} are found to be very useful in suppressing hadronic SM backgrounds in the boosted region. Because of the relatively large mass splitting between $\tilde{t}_2/\tilde{b}_1$ and $\tilde{t}_1$,  the $h/Z/W$ bosons from the heavier squarks decay usually are well boosted. Considering the $\tilde{t}_2 \to Z \tilde{t}_1$ process as an example and taking $m_{\tilde{t}_1}=500$ GeV,  we plot parton level distributions of the transverse momentum of $Z$ boson and the angular distance between two fermions from $Z$ decay in Fig.~\ref{pkins}. We can see from the figure that the typical transverse momentum of $Z$ boson exceeds $\sim 150 ~(200)$ GeV while the angular distance between the $Z$ boson decay products $\Delta R(f,f)$  which is roughly proportional to  ${2 m_Z}/{p_T(Z)}$ typically is less than $\sim$1.5 (1.0) for $m_{\tilde{t}_2}=800 ~(1000)$ GeV.

The closeness of $Z$ boson decay products indicates that they can be reconstructed as a whole, i.e., boson jet. 
A boson jet which has high invariant mass and appropriate substructure can be distinguished from QCD jet, thus provided as a most important handle for searching our benchmark points. Besides, there will be extra activities from the subsequent lighter stop $\tilde{t}_1$ decay such as leptons and b-jets. In the following, we propose a search for final state with two boson jets alongside with extra leptons/b-jets. 

\subsection{Signal and background analysis}

The signal processes that we are aiming to search for are
\begin{align}
p ~p &\to \tilde{t}_2 ~\bar{\tilde{t}}_2, ~ \tilde{t}_2 \to h/Z  ~\tilde{t_1} ~,~ \\
p~ p & \to \tilde{b}_1~ \bar{\tilde{b}}_1, ~ \tilde{b}_1 \to W ~\tilde{t_1} ~,~
\end{align}
with decay branching fractions of $\tilde{t}_2$, $\tilde{b}_1$ and $\tilde{t_1}$ given in the Tab.~\ref{bps}. 
At the LHC, the signal events can be trigged by requiring large missing transverse momentum in the final state, $\slashed{E}_T > 200$ GeV. As for event reconstruction, we first identify isolated electrons and muons with $p_T(e,\mu)>10$ GeV and $|\eta(e,\mu)|<2.5$, where the isolation means the scalar sum of transverse momenta of all particles with $p_{T}>0.5$ GeV that lie within a cone of radius $R=0.5$ around the $e(\mu)$ is less than 12\%(25\%) of the transverse momentum of $e(\mu)$.
Next, tracks that not belong to isolated leptons as well as neutral particles are used for jet clustering with fastjet~\cite{Cacciari:2011ma}. 
We adopt the BDRS method~\cite{Butterworth:2008iy} for tagging boosted boson jets: 
(1) reconstructing the boson jet candidates (fat jet) using C/A algorithm~\cite{Dokshitzer:1997in} with radius R=1.2 and $p_T >150$ GeV; 
(2) breaking each fat jet by undoing the clustering procedure. The two boson jets ($V_1, ~V_2$) are taken as the two leading fat jets with highest transverse momenta that have large mass drop $\mu<0.67$ and not too asymmetric mass splitting $y>0.09$ at any step during the declustering; 
(3) filtering each of the boson jets neighbourhood by reruning the C/A algorithm with a finer angle $R_{filt} = \min(0.3,R_{j_1,j_2}/2)$ and taking the three hardest subjets; 
(4) applying b-tag on the two leading subjets, where we have followed the b-tagging method that is used in Delphes: identifying the hadronic jet as the generated quark with largest PDG number that lies within the distance of $\Delta R < R_{filt}$ of the jet axis. The probabilities of b-tagging a b-jet, c-jet and light flavor jet are taken as 0.7, 0.2 and 0.005 respectively~\cite{Aad:2015ydr}. 
At last, for event that contains two boson jet candidates, we proceed further to reconstruct narrow jets. The constituents of the two boson jet candidates are removed from particle-flow objects of Delphes output. The remnants are clustered using the anti-$k_T$ jet clustering algorithm~\cite{Cacciari:2008gp} with jet cone radius of R = 0.4 and $p_T(j) > 20$ GeV to form narrow jets. The b-tagging is applied to each of the narrow jets with $|\eta(j)|<2.5$.   
During the reconstruction, the signal events are required to pass two more preselection cuts: the transverse momenta of two boson jets $p_T(V_1), p_T(V_2) > 200$ GeV and two boson jets should contain either no b-tagged subjet or exactly two b-tagged subjets.  

\begin{table}[htb]
\begin{center}
\begin{tabular}{|c|c|c|c|c||c|c|c|c|} \hline
 & \multicolumn{4} {|c|} { $m_{\tilde{t}_2/ \tilde{b}_1} \sim 800$ GeV} &  \multicolumn{4} {|c|} { $m_{\tilde{t}_2/ \tilde{b}_1} \sim 1000$ GeV}   \\
  & T1BC & T14B & T1BW & T1TN & T1BC & T14B & T1BW & T1TN \\ \hline \hline
$\sigma(\tilde{t}_2\bar{\tilde{t}}_2, \tilde{b}_1\bar{\tilde{b}}_1)$ (NLO) / fb & 99.6 & 76.5 & 76.5 & 76.5 & 23.0 & 24.5 & 24.5 & 24.5    \\ \hline
$\epsilon^{\text{pre}} \times \sigma$ /fb& 6.75 & 7.47 & 8.23 & 7.95 & 3.02 & 4.71 & 4.96 & 4.76  \\ \hline
\end{tabular}
\caption{\label{xsecsigs} Cross sections of benchmark points before and after preselections at 14 TeV LHC. }
\end{center}
\end{table}

The cross sections of benchmark points at 14 TeV LHC before and after the preselection are given in Tab.~\ref{xsecsigs}. The Next-to-Leading-Order (NLO) production cross section of $\tilde{t}_2\bar{\tilde{t}}_2$ plus $\tilde{b}_1\bar{\tilde{b}}_1$ are calculated by Prospino2~\cite{Beenakker:1997ut}. It can be seen that the signal rate decrease dramatically for increasing the particle mass. The preselection efficiencies is around 10\% for benchmark points with $m_{Q_3}=800$ GeV and become two times larger when $m_{Q_3} = 1$ TeV. 

We list all possible SM backgrounds for our signal in Tab.\ref{tab:smbg}, as well as their higher order production cross section at the LHC. After the preselection, the dominant backgrounds are $t\bar{t}$, diboson $+$ jets and $tW$ processes, in which either an energetic top quark or a QCD jet will be mis-tagged as a boson jet in our analysis, and the large missing transverse momentum is mainly due to the existence of neutrino in the final state. 

\begin{table}[h]
 \begin{center}
  \begin{tabular}{|c|c|c|c|} \hline
  Process & Total cross section & $\epsilon^{\text{pre}} \times \sigma$ \\ \hline
  $t\bar{t}$ & 953.6 pb (NNLO)~\cite{Czakon:2013goa}  & 252.3 fb \\ \hline
  $t\bar{t}Z$ & 1.12 pb (NLO)~\cite{Kardos:2011na}   &  \multirow{2}{*}{6.97 fb}\\
  $t\bar{t} W$ & 769 fb (NLO)~\cite{Campbell:2012dh}  &\\ \hline
  $t\bar{t} h$ & 604 fb (NLO)~\cite{Heinemeyer:2013tqa} & 1.66 fb \\ \hline
  $tW$(+j)  & 83.6 pb (NNLO)~\cite{Kidonakis:2013zqa}   & 41.5 fb \\ \hline
  $WW$(+2j) &126 pb(NLO)~\cite{Campbell:2011bn}  & \multirow{3}{*}{203.8 fb} \\
  $WZ$(+2j) & 31.9+20.3 pb (NLO)~\cite{Campbell:2011bn}  & \\
  $ZZ$(+2j) & 17.7 pb (NLO)~\cite{Campbell:2011bn}  &\\ \hline
  $Wh$(+2j) &  951+606 fb (NLO)~\cite{Campbell:2016jau}  & \multirow{2}{*}{1.04 fb} \\
  $Zh$(+2j) & 880 fb (NLO)~\cite{Campbell:2016jau}  & \\   \hline
  \end{tabular}
 \end{center}
   \caption{\label{tab:smbg}  Cross sections of backgrounds before and after preselections at 14 TeV LHC.}
\end{table}

Comparing Tab.~\ref{xsecsigs} and Tab.~\ref{tab:smbg}, we can find the production rates of our signals are around two order of magnitude smaller than that of backgrounds after the preselection cuts. Even at the 14 TeV LHC with integrated luminosity of 100 fb$^{-1}$, the signal significances are only around two.  Moreover, because of the smallness of signal-to-background ratios, the results are quite vulnerable to the systematic uncertainty. 
We need to apply more refined cuts to obtain higher signal significance as well as signal-to-background ratio. 

\begin{figure}[th]
\begin{center}
\includegraphics[width=0.47\textwidth]{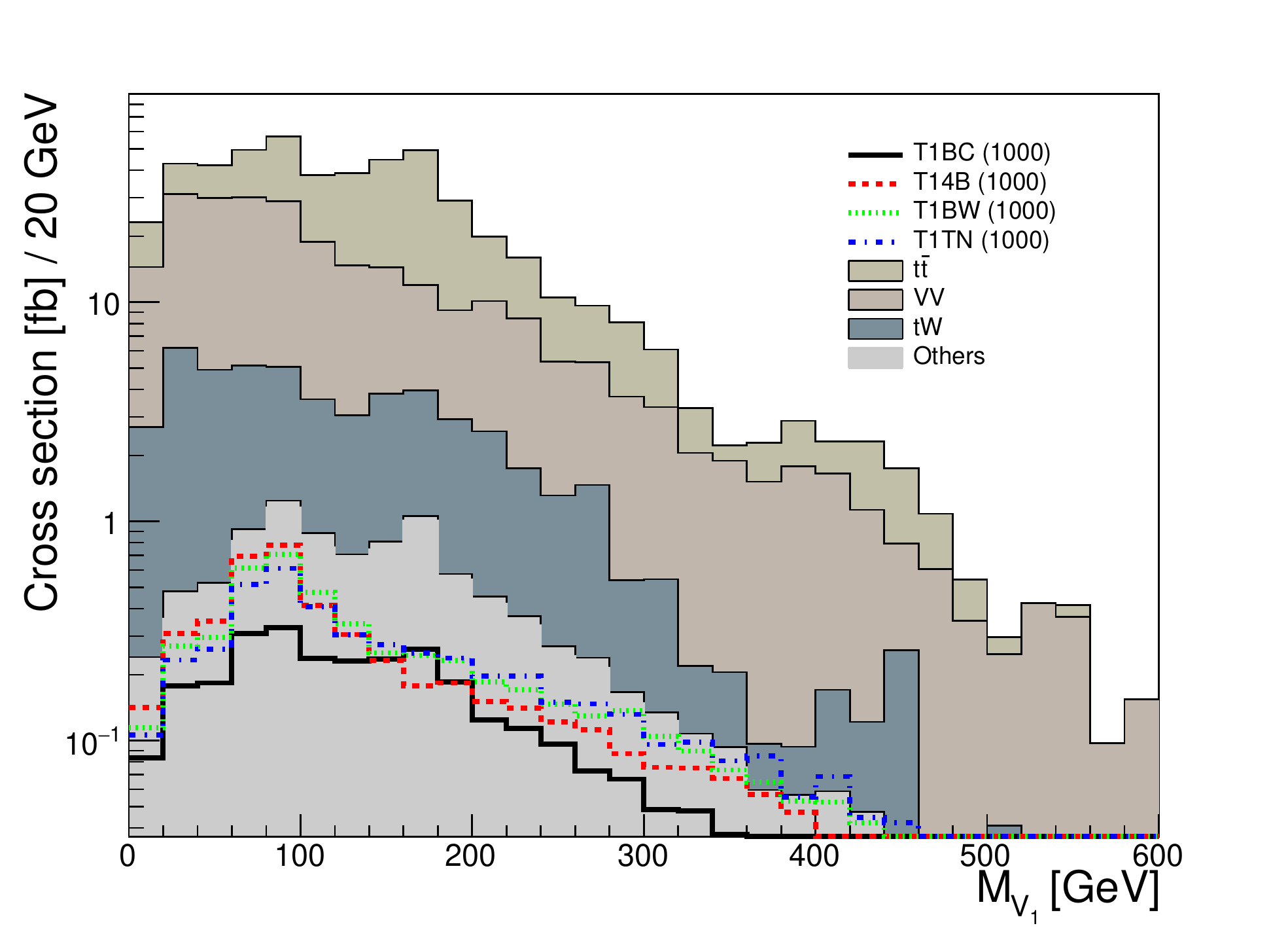}
\includegraphics[width=0.47\textwidth]{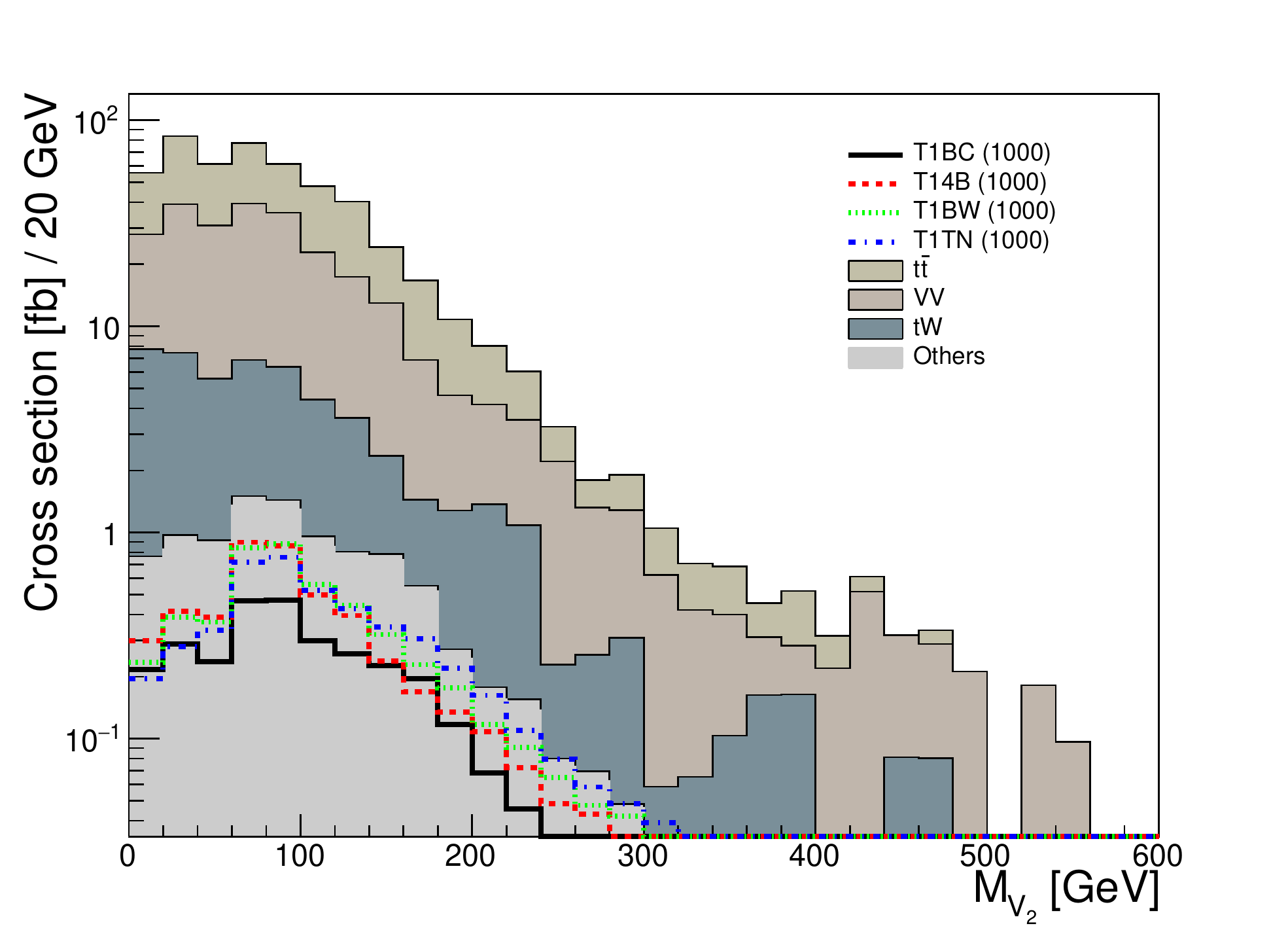} \\
\includegraphics[width=0.47\textwidth]{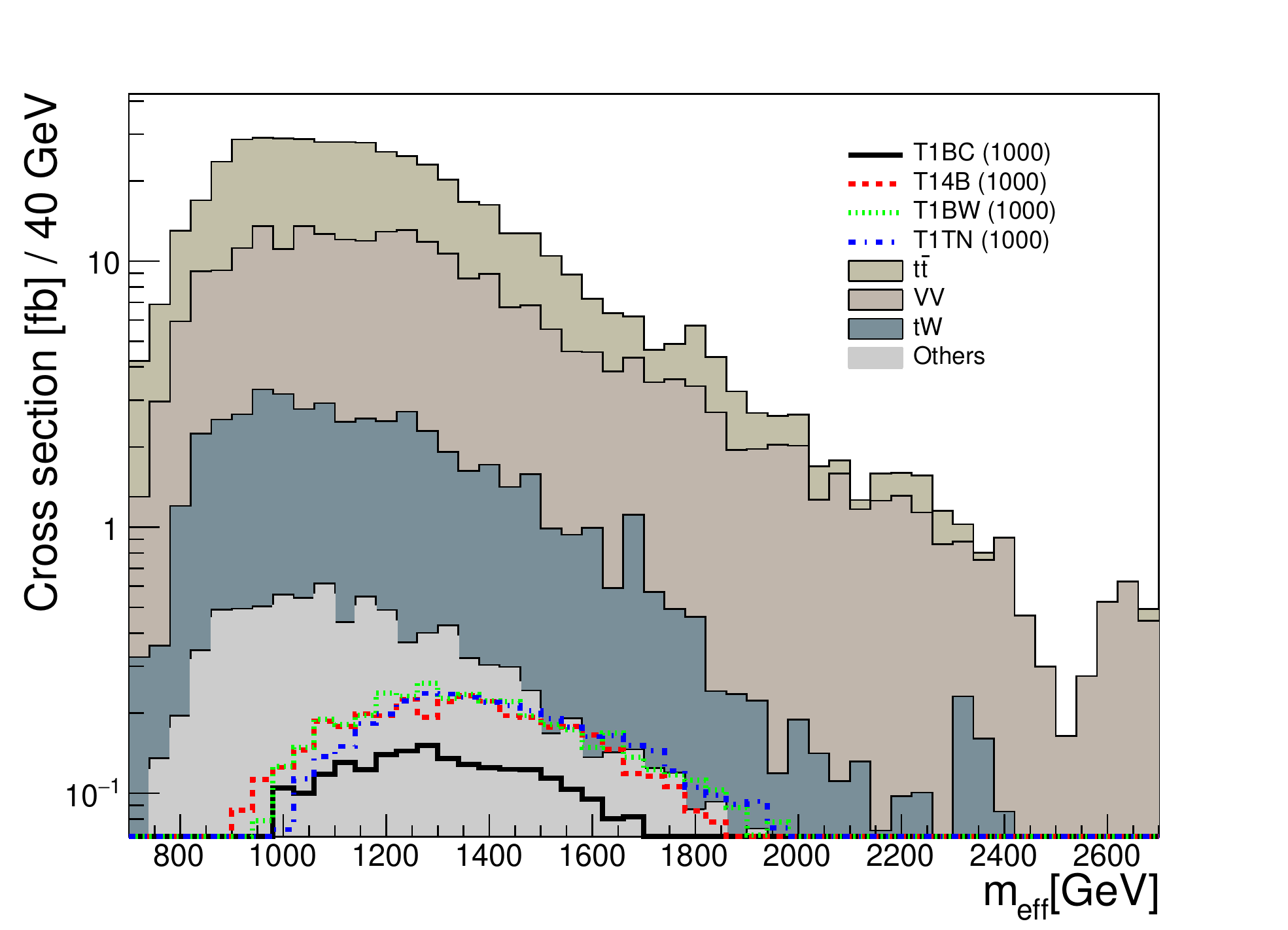}
\includegraphics[width=0.47\textwidth]{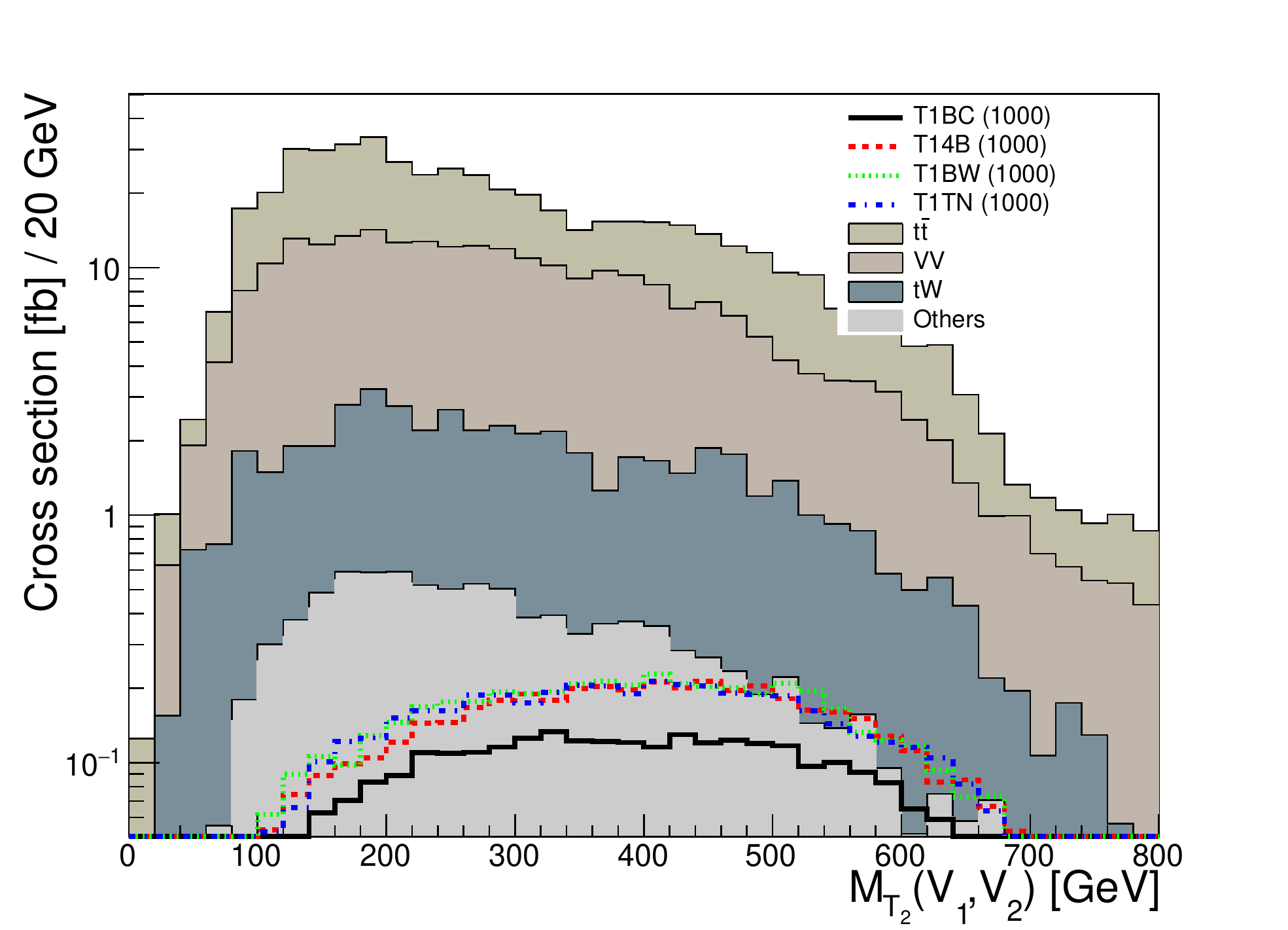}
\end{center}
\caption{\label{kins} Kinematic distributions for signals and backgrounds after preselection.  }
\end{figure}

\begin{figure}[th]
\begin{center}
\includegraphics[width=0.47\textwidth]{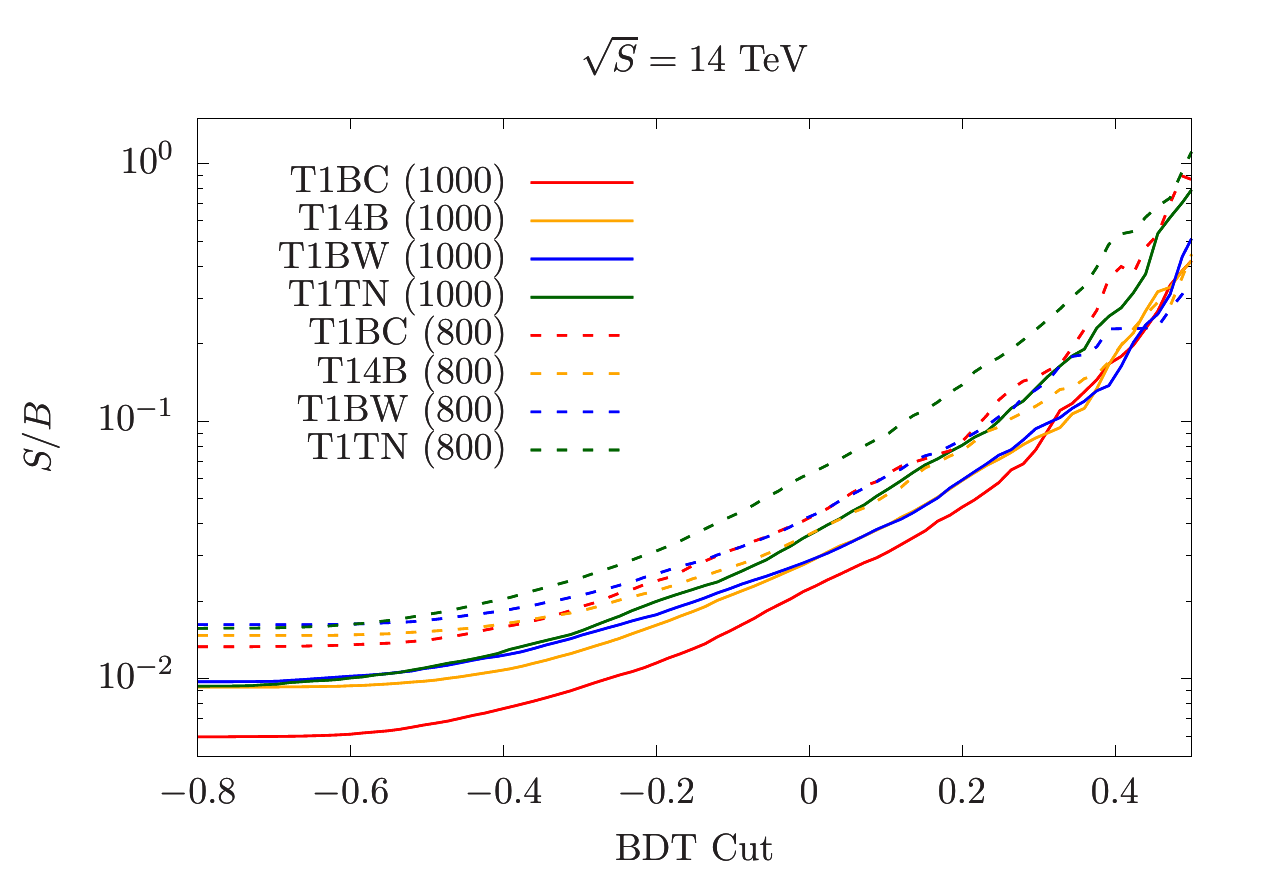}
\includegraphics[width=0.47\textwidth]{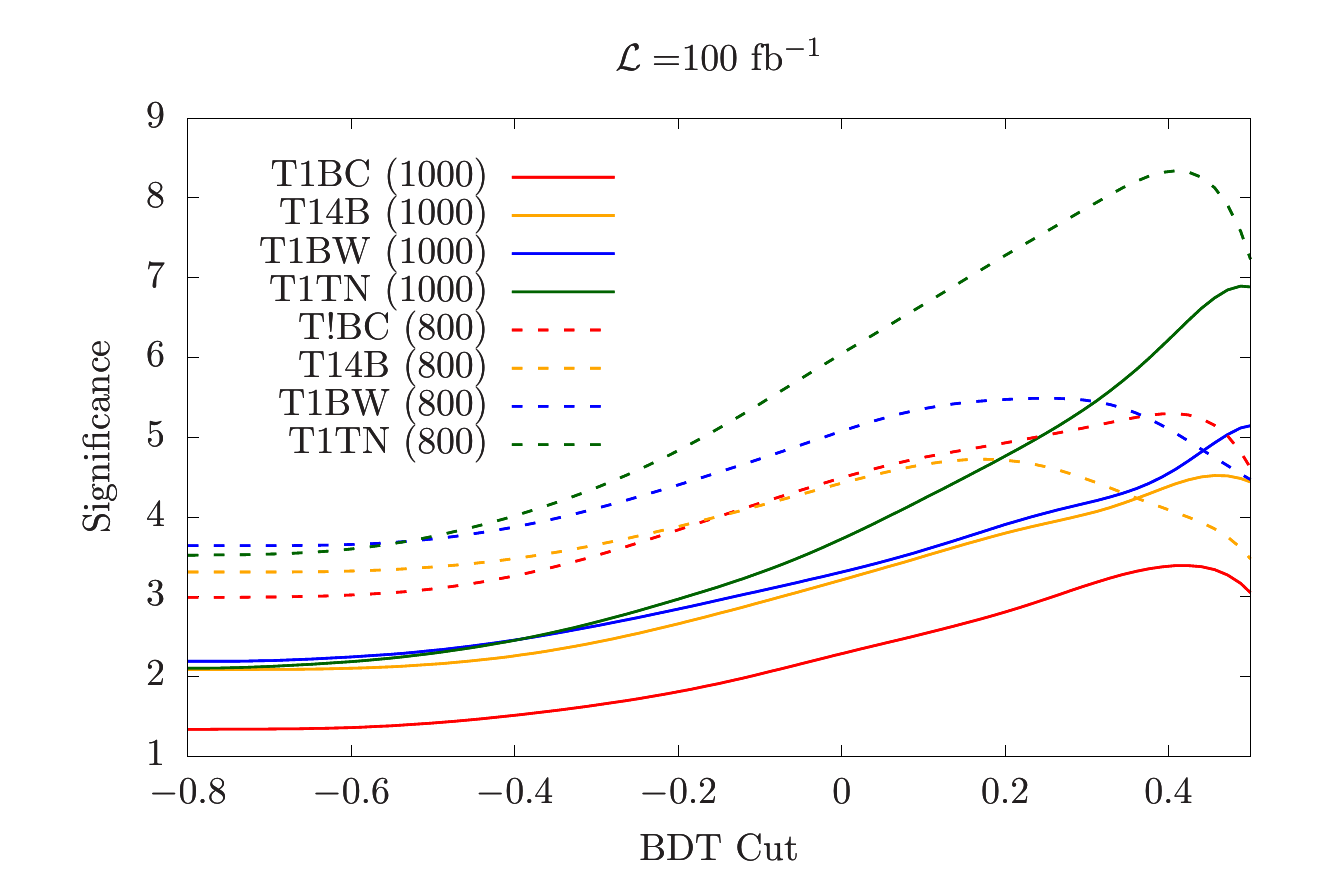}
\end{center}
\caption{\label{sigs} Left: the signal-background ratio for varying  BDT cut. Right: the signal significance at integrated luminosity of 100 fb$^{-1}$ for varying BDT cut. }
\end{figure}

First of all, the invariant masses of two boson jets should be close to either of $W/Z/h$ masses in signal processes. In the top panels of Fig.~\ref{sigs}, we plot the distributions of the invariant mass of boson jets ($m_{V_1}$, $m_{V_2}$) after pruning~\cite{Ellis:2009me}~\footnote{The BDRS method uses the filtered mass for the boson jet. But we find the distribution of pruned mass has a sharper peak in our case~\cite{Aad:2013gja}. }. 
In the figure, all backgrounds have been stacked up with contribution of each process indicated by different colors and distributions have been normalized to their production cross sections at 14 TeV LHC. We can see that most of signal events have $m_{V_1}$ and $m_{V_2}$ falling between [60,100] GeV since the branching ratio to $h$ is suppressed. While the backgrounds have relatively flat distribution between [20,200] GeV, especially for the $m_{V_1}$. This is because of the mis-tagging of top quark that enhanced the background rate at  $m_{V_1}\sim m_t$. It has to be noted that for the benchmark point T1BC the Br$(\tilde{t}_2 \to t \tilde{\chi}^0_i)$ and Br$(\tilde{b}_1 \to t \tilde{\chi}^\pm)$ are also sizeable. This leads to enhanced event rate at $m_{V_1} \sim m_t$ as well. 

The effective mass for our signal processes
\begin{align}
m_{\text{eff}} =  \slashed{E}_T +p_T(V_1) +p_T(V_2) +\sum p_T(\ell) +\sum p_T(j),
\end{align}
which is correlated with $m_{\tilde{t}_2 / \tilde{b}_1}$, should be higher than that for background processes.  As shown in the lower-left panel of Fig.~\ref{sigs}, the preselection render the $m_{\text{eff}}$ distribution of background peaks in a wide range between [1000,1200] GeV, while there are large fraction of signal events that have $m_{\text{eff}} >1200$ GeV.

Another useful discriminator that is used frequently in searching supersymmetry is the stransverse mass $M_{T_2}$~\cite{Lester:1999tx,Barr:2003rg}, which could reflect the mass difference between the squark and neutralino in the squark pair production channel with subsequent two body decay $\tilde{q} \to q \tilde{\chi}^0$. By drawing an analogy between our signal process $\tilde{t}_2 / \tilde{b} \to V \tilde{t}_1$ and $\tilde{q} \to q \tilde{\chi}^0$, we can define the modified stransverse mass as 
\begin{align}
M_{T_2} (V_1, V_2) = \min_{\vec{p}^1_T +\vec{p}^2_T = \vec{\slashed{p}}_T + \sum \vec{p}_T(\ell) +\sum \vec{p}_T(j)} [ \max (m_T(\vec{p}(V_1),\vec{p}^1_T ) ,m_T(\vec{p}(V_2),\vec{p}^2_T ) )],
\end{align}
where the $\sum \vec{p}_T(\ell)$ and $\sum \vec{p}_T(j)$ are vector sum of the transverse momenta of isolated leptons and narrow jets. The $M_{T_2} (V_1, V_2)$ distribution for signals and backgrounds are presented in the low-right panel of Fig.~\ref{kins}. We can see the signal events are occupying at larger values of $M_{T_2} (V_1, V_2)$ than backgrounds events. 

In order to obtain better signal and background discrimination, we employ the BDT method that takes into account the distribution profiles of the following variables 
\begin{align}
p_T(V_1), ~m_{V_1}, ~p_T(V_2), ~m_{V_2}, ~\slashed{E}_T, ~m_{\text{eff}}, ~M_{T_2} (V_1 ,V_2). 
\end{align}
Furthermore, the information from the decay products of the light top squark may help to improve our signal identification. So we consider three more variables in the BDT analysis: 
\begin{align}
n_{\ell}, ~n_b, ~p_T(\ell_1),
\end{align}
where the $p_T(\ell_1)$ is the transverse momentum of the leading lepton if it exists. 

The BDT method uses a 100 tree ensemble that requires a minimum training events in each leaf node of 2.5\% and a maximum tree depth of two. It is trained on the half of the preselected signal and backgrounds events and is tested on the rest of the events. We also require the Kolmogorov-Smirnov test of the BDT analysis should be greater than 0.01 to avoid overtraining. 

Having the BDT response distributions for both signal and background, we can impose a cut on the BDT responses to improve the signal significance. 
Fig.~\ref{sigs} shows the signal-to-background ratios (left panel) and the signal significances with 100 fb$^{-1}$ data sample (right) for all benchmark points. The signal significance is calculated by
\begin{align}
\mathcal{S} = \sqrt{2((s+b) \ln(1+\frac{s}{b})-s)}. 
\end{align}
We can see that a cut of BDT$\gtrsim 0.3$ will maximize the signal significance and keep the signal-to-background ratio at $\mathcal{O}(10)$\% level. 

\begin{figure}[th]
\begin{center}
\includegraphics[width=0.8\textwidth]{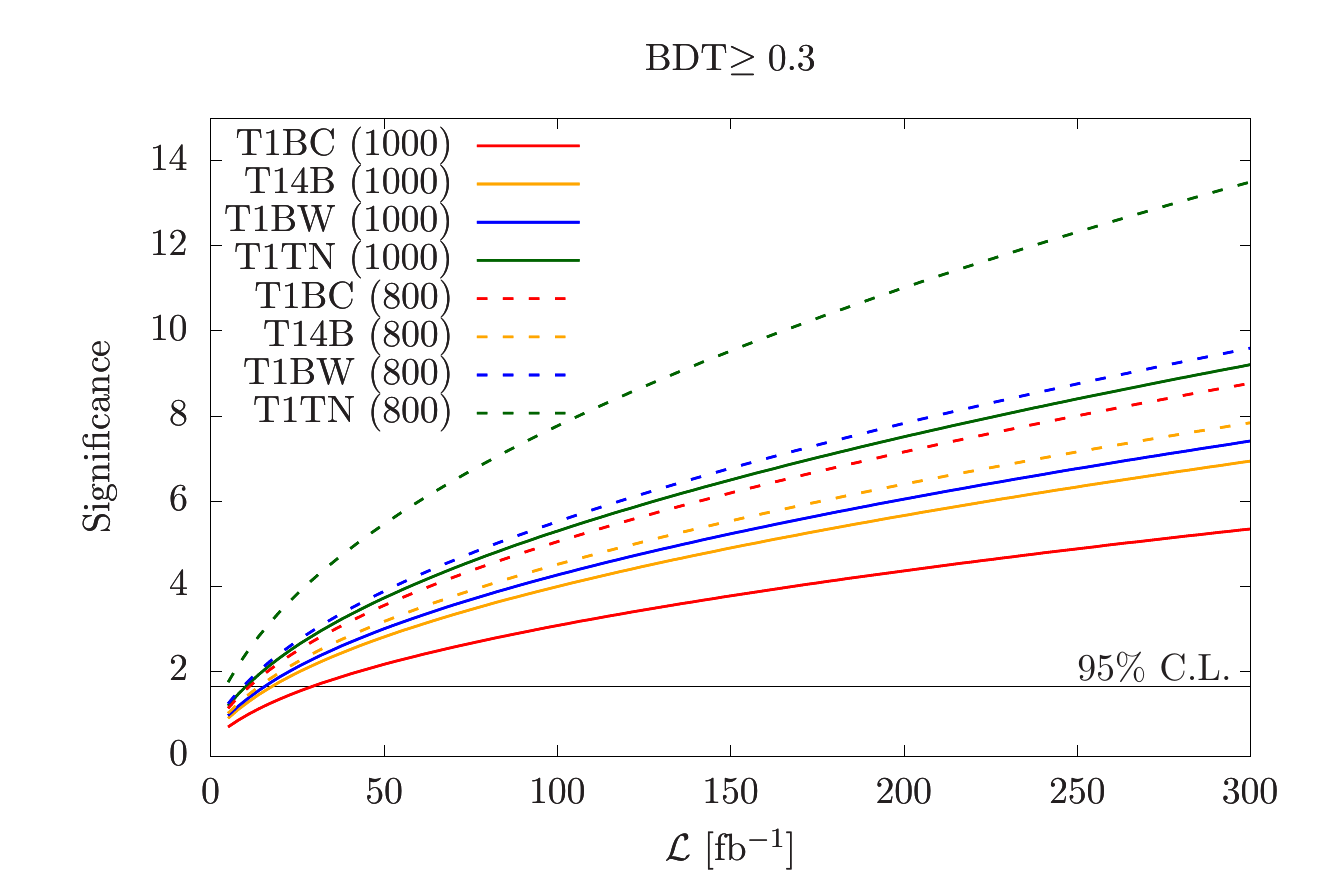}
\end{center}
\caption{\label{signal-to-background ratio} The signal significance at different integrated luminosities, where the BDT cut is chosen as BDT$\ge 0.3$. The horizontal line corresponds to the 95\% C.L. exclusion limit. }
\end{figure}

In Fig.~\ref{signal-to-background ratio}, we plot the signal significances for all benchmark points with different integrated luminosity, where we have chosen the cut BDT$\ge 0.3$. A heavier stop sector of $\sim 1$ TeV can be excluded at 95\% C.L. at very early stage of the LHC run-II. Since the lighter stops $\tilde{t}_1$ of benchmark points are way beyond the reach of the LHC search at 13 TeV 13.3 fb$^{-1}$, we can conclude that the heavier stop provide a better chance for searching supersymmetry.  
Moreover, comparing to the method in Ref.~\cite{Cheng:2016npb,Pierce:2016nwg} which utilise the leptons and b-jets in the final state, our search strategy can achieve a few times larger signal significance because of higher signal rate. 

\section{Conclusion}

A quasi-natural pattern of low energy supersymmetry, in which the lighter stop has mass around a few hundred GeV and to be close to the LSP mass while the heavier stop and the lighter sbottom have masses around TeV is considered in this work. 
In this scenario, due to the compressed mass between the lighter stop and the LSP,  the lighter stop decay can only produce soft leptons/jets in the final state; thus it evades all current LHC searches and is difficult to probe in future experiments. 
The heavier stop $\wt t_2$ and lighter sbottom $\wt b_1$, in contrast, may provide a better handle for searching the compressed SUSY. 

In the framework of MSSM, considering either the bino or the Higgsino as the LSP, we find that the bosonic modes $h/Z \tilde{t}_1$ ($W \tilde{t}_1$) dominate the $\tilde{t}_2$ ($\tilde{b}_1$) decay in the parameter space with relatively large left-right stop mixing as well as large trilinear coupling $A_t$. With a moderately large mass gap between the heavier members and lightest stop, the bosons in the decay chain are generically quite energetic. This allows us to employ the jet substructure technique for discriminating the natural SUSY signals in searches at the LHC. 

We consider the discovery prospects of eight benchmark points at the LHC-14, in terms of four possible decay modes of the lighter stop $\tilde{t}_1$: (1) $\tilde{t}_1 \to b \tilde{\chi}^\pm_1$; (2) $\tilde{t}_1 \to b f f \tilde{\chi}^0_1$; (3) $\tilde{t}_1 \to b W \tilde{\chi}^0_1$; (4) $\tilde{t}_1 \to t \tilde{\chi}^0_1$ as well as two different masses of $\tilde{t}_2/ \tilde{b}_1$: (a) $m_{\tilde{t}_2/ \tilde{b}_1} \sim 800$ GeV; (b) $m_{\tilde{t}_2/ \tilde{b}_1} \sim 1000$ GeV. 
We search for the $\tilde{t}_2 \bar{\tilde{t}}_2$ and $\tilde{b}_1 \bar{\tilde{b}}_1$ production in final state with two boosted boson jets that have substructures and high invariant masses, leptons/b-jets and MET. After considering background contamination and adopting the BDT method for signal discrimination, we find that a heavier stop and lighter sbottom with masses $\sim 1$ TeV can be excluded at 95\% C.L. with integrated luminosity of 10-30 fb$^{-1}$.   
Among the four decay modes of the $\tilde{t}_1$, the search sensitivities decrease from $t\tilde{\chi}^0_1$ to $bW \tilde{\chi}^0_1$ to $b ff \tilde{\chi}^0_1$, as the mass difference between the $\tilde{t}_1$ and $\tilde{\chi}^0_1$ is successively smaller. The $b\tilde{\chi}^\pm_1$ mode has the least search sensitivity. This is because this mode is possible only when Higgsino is the LSP. Then the decay branching ratios of $\tilde{t}_2 \to t \tilde{\chi}^0$ / $\tilde{b}_1 \to t \tilde{\chi}^\pm$  become competitive to that of the bosonic decay of $\tilde{t}_2$ / $\tilde{b}_1$. 
Finally, we note that with the aid of the jet substructure and BDT analysis, our search strategy can achieve a few times larger signal significance than the searches proposed in Refs.~\cite{Cheng:2016npb,Pierce:2016nwg} which utilize the multiple leptons and b-jets in the final state.

\acknowledgments
This work is supported in part by National Research Foundation of Korea (NRF) Research Grant NRF-2015R1A2A1A05001869 (JL) and IBS under the project code IBS-R018-D1 (MZ).

\bibliographystyle{jhep}
\bibliography{stop2}

\end{document}